\documentclass[10pt,twocolumn,letterpaper]{article}

\usepackage{cvpr}              
\usepackage{siunitx}
\usepackage{multirow}
\usepackage{makecell}
\usepackage{booktabs,array}
\usepackage{balance}
\usepackage{wrapfig}
\usepackage{alphalph}
\usepackage{colortbl}

\usepackage{graphicx,tikz,libertine,xcolor}
\usetikzlibrary{calc,spy}

\usepackage{wrapfig}
\usepackage{enumitem}
\usepackage{adjustbox}
\usepackage{colortbl}

\usepackage{placeins}
\usepackage{acronym}
\usepackage{overpic}
\renewcommand*{\eqref}[1]{Eq.~\hyperref[#1]{\ref*{#1}}}
\newcommand*{\chref}[1]{Chapter~\hyperref[#1]{\ref*{#1}}}
\newcommand*{\secref}[1]{Sec.~\hyperref[#1]{\ref*{#1}}}
\newcommand*{\figref}[1]{Fig.~\hyperref[#1]{\ref*{#1}}}
\newcommand*{\figureref}[1]{Figure~\hyperref[#1]{\ref*{#1}}}
\newcommand*{\tabref}[1]{Tab.~\hyperref[#1]{\ref*{#1}}}
\newcommand*{\tableref}[1]{Table~\hyperref[#1]{\ref*{#1}}}
\newcommand*{\Subref}[1]{\hyperref[#1]{(\subref*{#1})}}

\newcommand{\recall}{\text{CD}_{\text{Inp}\rightarrow\text{Rec}}}
\newcommand{\precision}{\text{CD}_{\text{Rec}\rightarrow\text{Ref}}}

\newcommand{\tf}{\cellcolor{blue!25}}
\newcommand{\ts}{\cellcolor{blue!10}}
\newcommand{\rc}{\cellcolor{black!10}}

\makeatletter
\newcommand{\definetrim}[2]{%
  \define@key{Gin}{#1}[]{\setkeys{Gin}{trim=#2,clip}}%
}
\makeatother

\newcolumntype{H}{>{\setbox0=\hbox\bgroup}c<{\egroup}@{}}

\newcommand*{\affaddr}[1]{#1} 
\newcommand*{\affmark}[1][*]{\textsuperscript{#1}}

\definecolor{cvprblue}{rgb}{0.21,0.49,0.74}
\usepackage[pagebackref,breaklinks,colorlinks,citecolor=cvprblue]{hyperref}

\def\paperID{4} 
\def\confName{CVPR workshop on Urban Scene Modeling}
\def\confYear{2024}

\title{SimpliCity: Reconstructing Buildings with Simple Regularized 3D Models}

\author{%
Jean-Philippe Bauchet\affmark[1,*], Raphael Sulzer\affmark[1,2,*], Florent Lafarge\affmark[2], Yuliya Tarabalka\affmark[1]\\
\\
\affaddr{\affmark[1]LuxCarta Technology, Mouans-Sartoux, France}\\
\affaddr{\affmark[2]Centre INRIA d'Universit\'e C\^ote d'Azur, Sophia Antipolis, France}\\
}

\begin{document}
\twocolumn[{
    \renewcommand\twocolumn[1][]{#1}
    \maketitle
    \begin{center}
        \definetrim{mytrim1}{0 0 0 0}
\definetrim{mytrim2}{100 10 100 10}
\definetrim{mytrim3}{100 50 100 50}
\definetrim{mytrim4}{0 0 0 0}
\newcommand{\mywidth}{0.166\linewidth}
\newcommand{\mywidthb}{0.18\linewidth}
\newcommand{\myfontsize}{\scriptsize}

\setlength{\tabcolsep}{0mm}

\centering

\newcommand{\teaserwidth}{0.17\linewidth}

\tikzstyle{closeup} = [
  opacity=1.0,          
  height=1.0cm,         
  width=1.0cm,          
  connect spies, blue  
]
\tikzstyle{largewindow} = [circle, blue, line width=0.3mm]
\tikzstyle{smallwindow} = [circle, blue,line width=0.15mm]

\begin{tikzpicture}[x=29cm, y=5cm, spy using outlines={every spy on node/.append style={smallwindow}}]

  \coordinate (ai) at (-0.003,0.142);
  \coordinate (bi) at (-0.01,-0.06);

  \coordinate (aj) at (0.028,0);
  \coordinate (bj) at (-0.02,0);
  
  \coordinate (c) at (0,-0.15);
  
\node[anchor=south] (FigA) at (0,0) {\includegraphics[width=\teaserwidth]{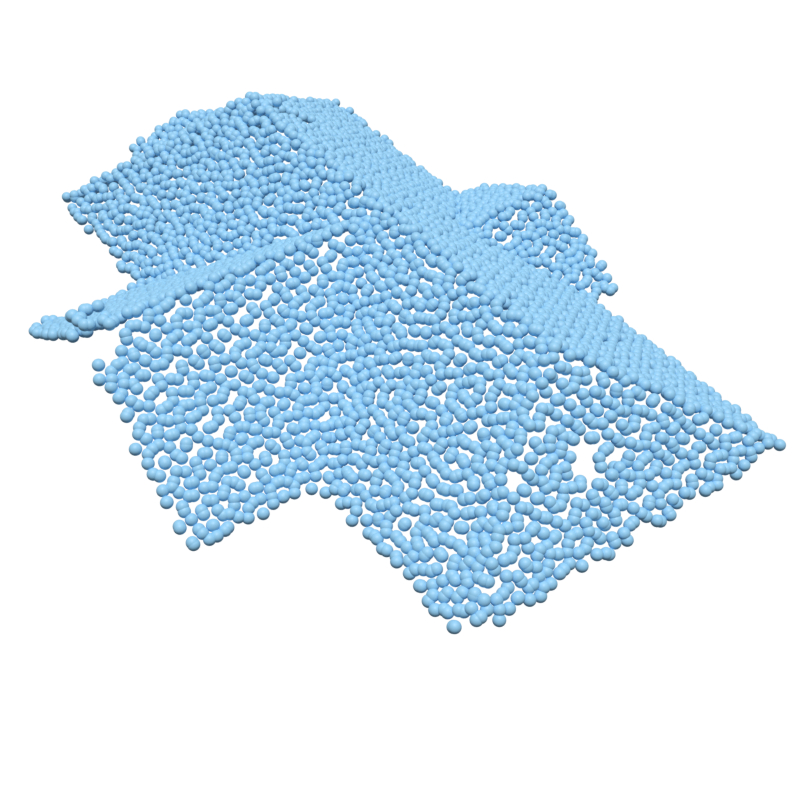}};
\spy [closeup,magnification=3] on ($(FigA)+(ai)$) 
    in node[largewindow] at ($(FigA.south)+(aj)$);
\spy [closeup,magnification=3] on ($(FigA)+(bi)$) 
    in node[largewindow] at ($(FigA.south)+(bj)$);
\node [anchor=north] at ($(FigA.south)+(c)$) {Input};
\node[anchor=south] (FigB) at (0.1,0) {\includegraphics[width=\teaserwidth]{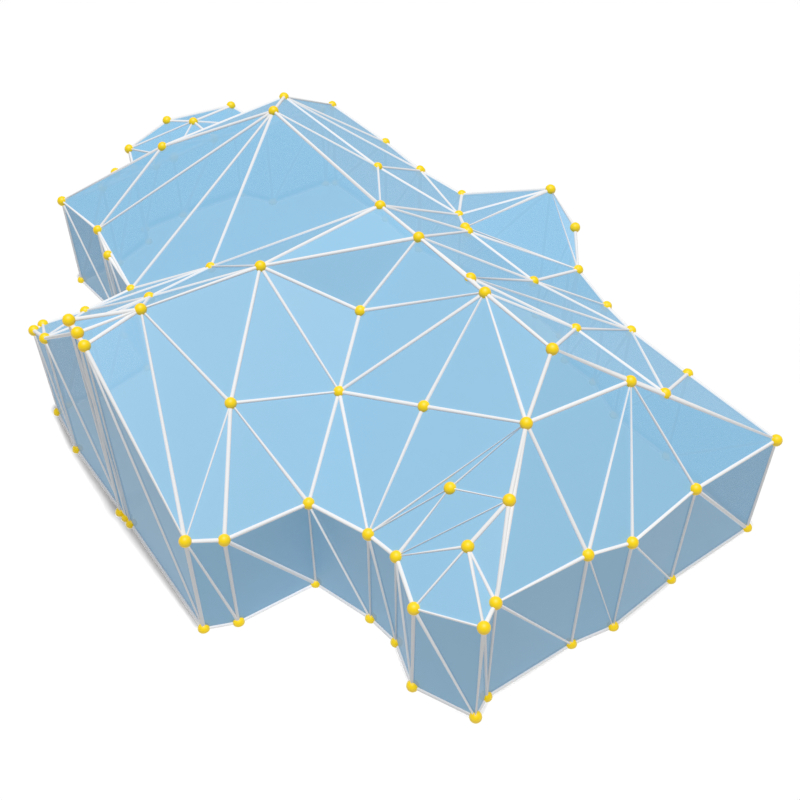}};
\spy [closeup,magnification=3] on ($(FigB)+(ai)$) 
    in node[largewindow] at ($(FigB.south)+(aj)$);
\spy [closeup,magnification=3] on ($(FigB)+(bi)$) 
    in node[largewindow] at ($(FigB.south)+(bj)$);
\node [anchor=north] at ($(FigB.south)+(c)$) {2.5DC~\cite{zhou2010dualcontouring}};
\node[anchor=south] (FigC) at (0.2,0) {\includegraphics[width=\teaserwidth]{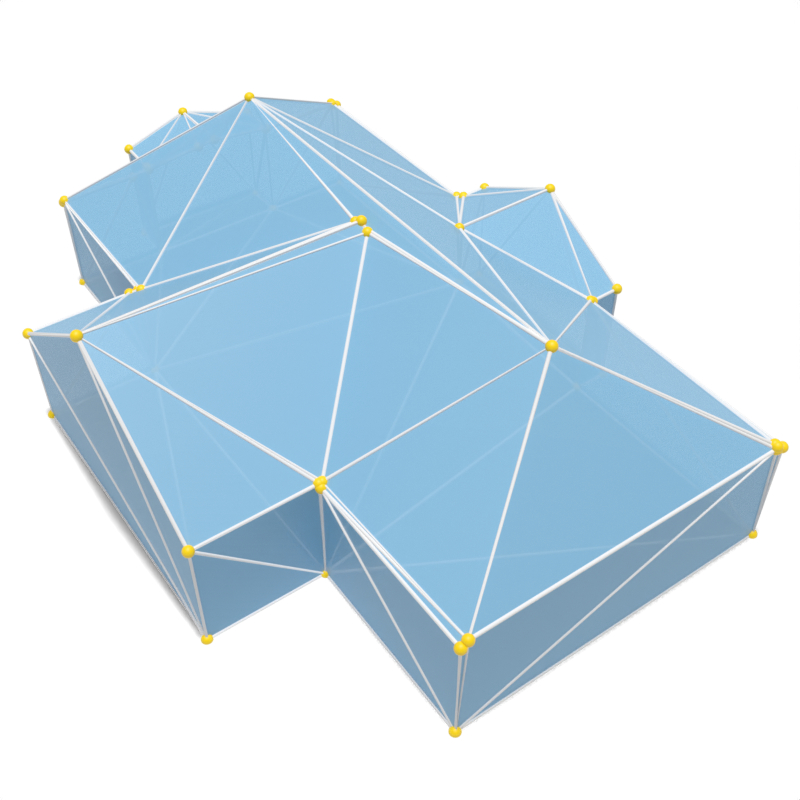}};
\spy [closeup,magnification=3] on ($(FigC)+(ai)$) 
    in node[largewindow] at ($(FigC.south)+(aj)$);
\spy [closeup,magnification=3] on ($(FigC)+(bi)$) 
    in node[largewindow] at ($(FigC.south)+(bj)$);
\node [anchor=north] at ($(FigC.south)+(c)$) {KSR~\cite{bauchet2020kinetic}};
\node[anchor=south] (FigD) at (0.3,0) {\includegraphics[width=\teaserwidth]{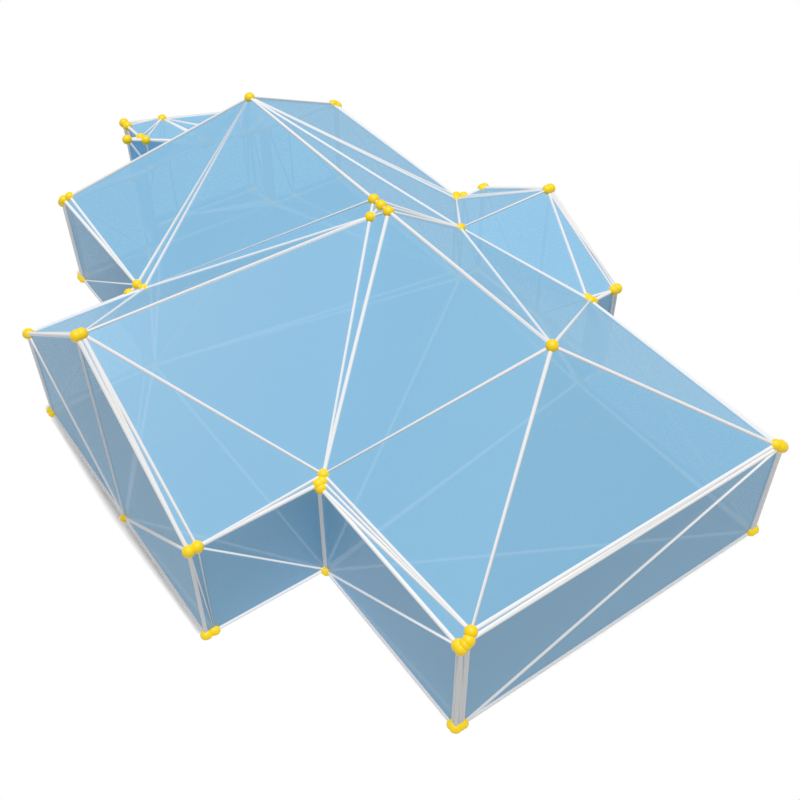}};
\spy [closeup,magnification=3] on ($(FigD)+(ai)$) 
    in node[largewindow] at ($(FigD.south)+(aj)$);
\spy [closeup,magnification=3] on ($(FigD)+(bi)$) 
    in node[largewindow] at ($(FigD.south)+(bj)$);
\node [anchor=north] at ($(FigD.south)+(c)$) {Geoflow~\cite{peters2022geoflow}};
\node[anchor=south] (FigE) at (0.4,0) {\includegraphics[width=\teaserwidth]{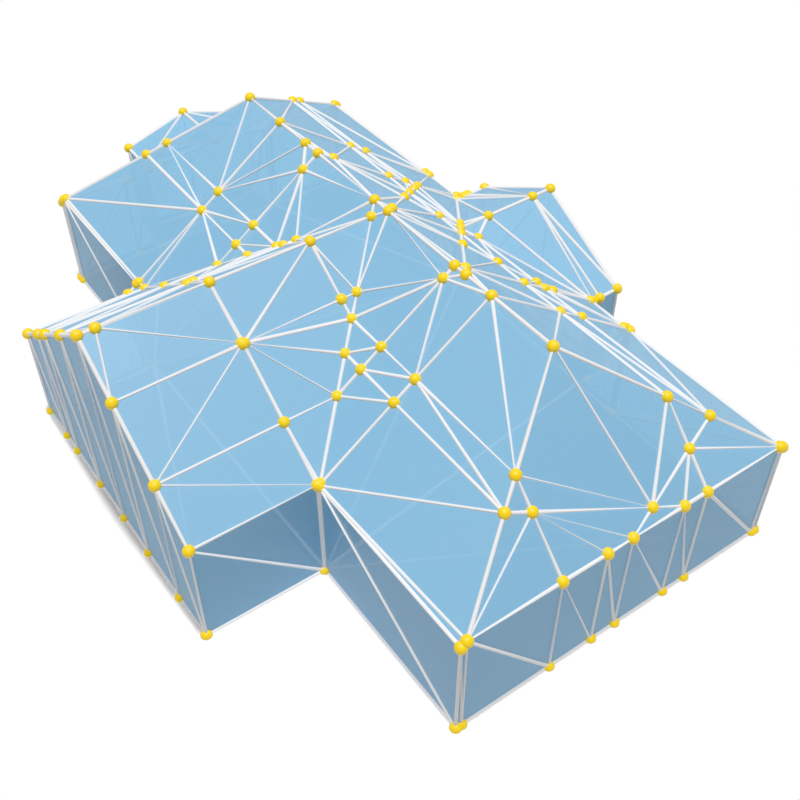}};
\spy [closeup,magnification=3] on ($(FigE)+(ai)$) 
    in node[largewindow] at ($(FigE.south)+(aj)$);
\spy [closeup,magnification=3] on ($(FigE)+(bi)$) 
    in node[largewindow] at ($(FigE.south)+(bj)$);
\node [anchor=north] at ($(FigE.south)+(c)$) {City3D~\cite{huang2022city3d}};
\node[anchor=south] (FigF) at (0.5,0) {\includegraphics[width=\teaserwidth]{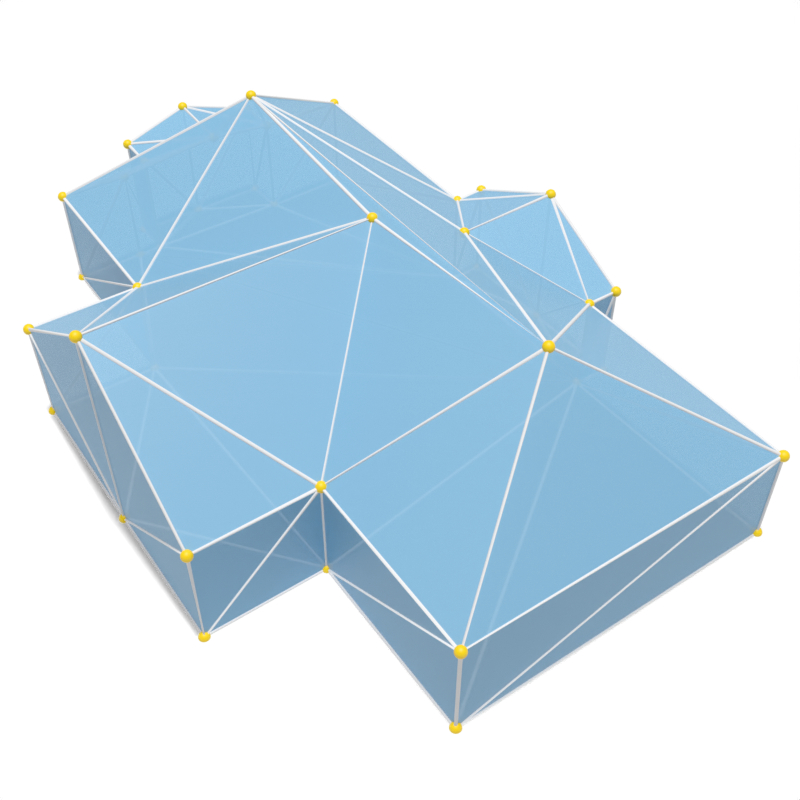}};
\spy [closeup,magnification=3] on ($(FigF)+(ai)$) 
    in node[largewindow] at ($(FigF.south)+(aj)$);
\spy [closeup,magnification=3] on ($(FigF)+(bi)$) 
    in node[largewindow] at ($(FigF.south)+(bj)$);
\node [anchor=north] at ($(FigF.south)+(c)$) {Ours};

\end{tikzpicture}

\captionof{figure}%
{\textbf{SimpliCity.} Our building reconstruction method produces a simple, regularized mesh while being a faithful approximation of the input Lidar scan. In contrast, the output meshes of Geoflow \cite{peters2022geoflow} and City3D \cite{huang2022city3d} contain a much higher number of facets, including tiny ones that correct connectivity approximations around corners.}
\label{fig:teaser}

    \end{center}
}]
\begin{abstract}
Automatic methods for reconstructing buildings from airborne LiDAR point clouds focus on producing accurate 3D models in a fast and scalable manner, but they overlook the problem of delivering simple and regularized models to practitioners. As a result, output meshes often suffer from connectivity approximations around corners with either the presence of multiple vertices and tiny facets, or the necessity to break the planarity constraint on roof sections and facade components. We propose a 2D planimetric arrangement-based framework to address this problem. 
We first regularize, not the 3D planes as commonly done in the literature, but a 2D polyhedral partition constructed from the planes. Second, we extrude this partition to 3D by an optimization process that guarantees the planarity of the roof sections as well as the preservation of the vertical discontinuities and horizontal rooftop edges.
We show the benefits of our approach against existing methods by producing simpler 3D models while offering a similar fidelity and efficiency.  
\end{abstract}

\newcommand{\customfootnotetext}[2]{{
  \renewcommand{\thefootnote}{#1}
  \footnotetext[0]{#2}}}
\customfootnotetext{*}{Both authors contributed equally to the paper.}

\section{Introduction}



Reconstructing buildings in 3D from physical data measurements is a long standing problem within the computer vision, remote sensing and computer graphics communities \cite{musialski2013survey,wang2018lidar}. While dense mesh-based representations are commonly used for visualization and immersive tasks, \eg GoogleEarth \cite{GoogleEarth}, lightweight 3D models of buildings are often required in many large-scale application scenarios such as simulation of physical phenomena, urban planning or navigation \cite{biljecki2015applications,deren2021smart}. In these scenarios, a building is represented by a surface mesh with a low number of facets, ideally just enough to describe its structure. 


Among possible data sources, airborne Lidar scanning offers several advantages. It directly produces 3D point clouds with a high precision and a density that can reach up to 50 points per squared meter.
The emitted pulses can also provide descriptions of buildings that are occluded by vegetation or shadows in optical imagery, making them hard to detect and reconstruct. Airborne Lidar scans used to have limited availability, but in a renewed interest for the construction of a digital twin, public organizations, such as the French National Mapping Agency \cite{ign} or the Swiss Federal Office of Topography \cite{swisstopo} have acquired and released Lidar data at a country scale. In addition, this public Lidar data is often already semantically classified \cite{gaydon2022myria3d}.


We address the problem of reconstructing lightweight building models from airborne Lidar point clouds with the following objectives:

\begin{itemize}
\item Fidelity: output meshes should constitute of a faithful approximation of the data,
\item Simplicity: output meshes should be composed of a low number of facets and preserve the geometric regularities of buildings, if any, 
\item Efficiency: the algorithm should be fast, scalable, and proceed with a low number of user parameters,
\item Geometric guarantees: the output meshes should be watertight, 2d-manifold, intersection-free, and should conform to the CityGML LOD2 level of detail \cite{groger2012citygml} with, in particular, the planarity of facets composing a facade component or a roof section. 
\end{itemize}

Unfortunately, existing methods do not perform well on these four objectives simultaneously. Robust approaches \cite{bauchet2019city,huang2022city3d,zebedin2008fusion,zhou_cvpr12} usually target high fidelity and efficiency by detecting planes and assembling them either directly in 3D or within a planimetric arrangement extruded to 3D. Although planes can be accurately and efficiently detected from airborne Lidar, plane-based approaches often generate output meshes that lack regularity, simplicity and important geometric guarantees. A recurring problem is that four or more planes are unlikely to be connected in exactly one vertex leading to the creation of tiny facets approximating the one-vertex connection, or the loss of the planarity property of polygonal facets (\cf Fig. \ref{fig:teaser}). Furthermore, geometric regularities are often poorly preserved in the output models. Regularizing detected planes \cite{zhou_cvpr12,verdie_tog15} can help but does not guarantee a regular output mesh free of geometric defects.  

In this work, we propose a 2D planimetric arrangement-based framework to address these issues. The two key ideas are first to regularize, not the 3D planes as in the literature, but the 2D polyhedral partition constructed from the planes, and second to extrude in 3D this partition by an optimization process that guarantees the planarity of the roof sections as well as the preservation of the vertical discontinuities and horizontal rooftop edges. Our solution is efficient in solving issues (i) and (ii) as the 2D polygonal partition exhibits a simple facet connectivity while being highly regularized. We show the benefits of our approach against prior methods by producing more simple 3D models while reaching a similar fidelity and efficiency.

\section{Related work}
\label{relatedwork}

Our review of previous work discusses the main strategies for reconstructing buildings from input point clouds. Note that a vast literature also exists from other data sources such as single-view satellite image \cite{mahmud2020boundary, zhang2022procedural, lussange20233d} or dense meshes generated by multiview stereo pipelines \cite{zhu2018large, han2021urban}.   


\vspace{-0.2cm}
\paragraph{Model-driven.} These methods typically perform template matching from a predefined library of building parts \cite{verma20063d,kada20093d,nys2020cityjson}. Roof topology graphs \cite{xiong2014graph, xiong2015flexible} or rules derived from constructive solid geometry  \cite{lafarge2010structural} can be used to guide the matching. 
Such methods produce realistic results, but lose generality when applied to urban landscapes that cannot be precisely described using the predefined templates. The Manhattan-world assumption can also be used to restrict the orientations of buildings to three orthogonal directions, and reconstruct them as polycubes \cite{vanegas2010building, li2016manhattan}.


\vspace{-0.2cm}
\paragraph{Mesh simplification.} Other methods simplify a dense mesh reconstructed from the input point cloud \cite{kazhdan2013screened, boulch2022poco, sulzer2023survey} into a more concise mesh. A common strategy consists in iteratively collapsing edges depending on quadric error metrics \cite{garland1997surface}. To better preserve the structure of piece-wise-planar objects like buildings, these edge contraction operators may consider planar proxies detected during a preprocessing stage \cite{salinas2015structure, li2021feature, wang2021topology}. 
Closely related to these simplification techniques, dual contouring meshing techniques can also be used for reconstruct buildings with 2.5D-view dependent triangle meshes \cite{zhou2010dualcontouring, zhou201125dbuilding}. 


\begin{figure*}[t]
    \captionsetup[figure]{position=auto}
    \newcommand{\mywidth}{0.25\linewidth}
    \newcommand{\mywidthd}{0.38\linewidth}
    \definetrim{mytrim}{70 0 70 0}
    \definetrim{mytrimd}{20 70 40 70}

    \tikzstyle{closeup} = [
  opacity=1.0,          
  height=1.0cm,         
  width=1.0cm,          
  connect spies, blue  
]
\tikzstyle{largewindow} = [circle, blue, line width=0.3mm]
\tikzstyle{smallwindow} = [circle, blue,line width=0.15mm]

    \begin{tabular}{@{}c@{}c@{}c@{}c@{}}
    \centering


    \begin{subfigure}{\mywidth}
        \begin{overpic}[width=\linewidth,mytrim]
            {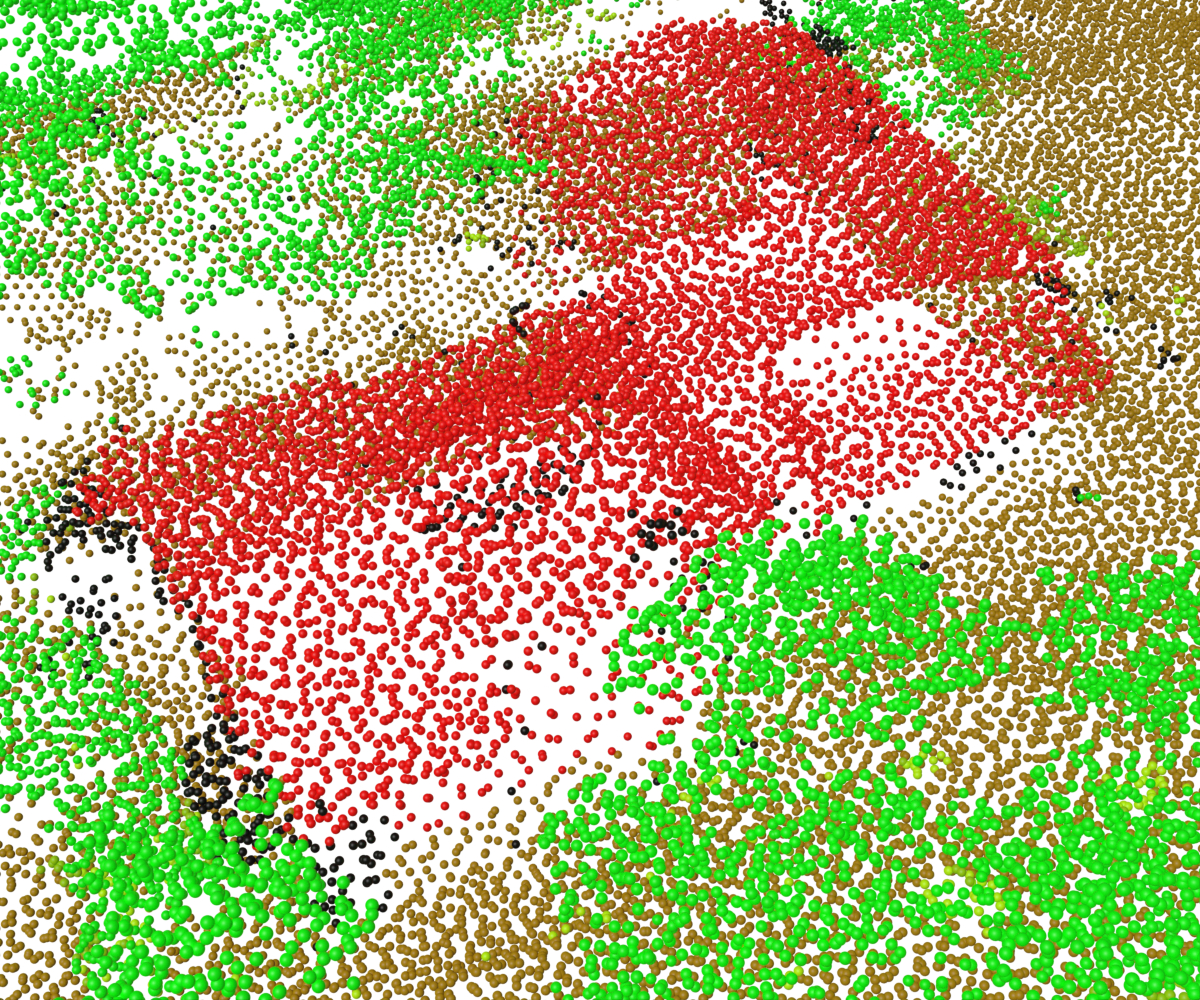}
            \put(68,0){%
                \includegraphics[scale=.04,mytrim]{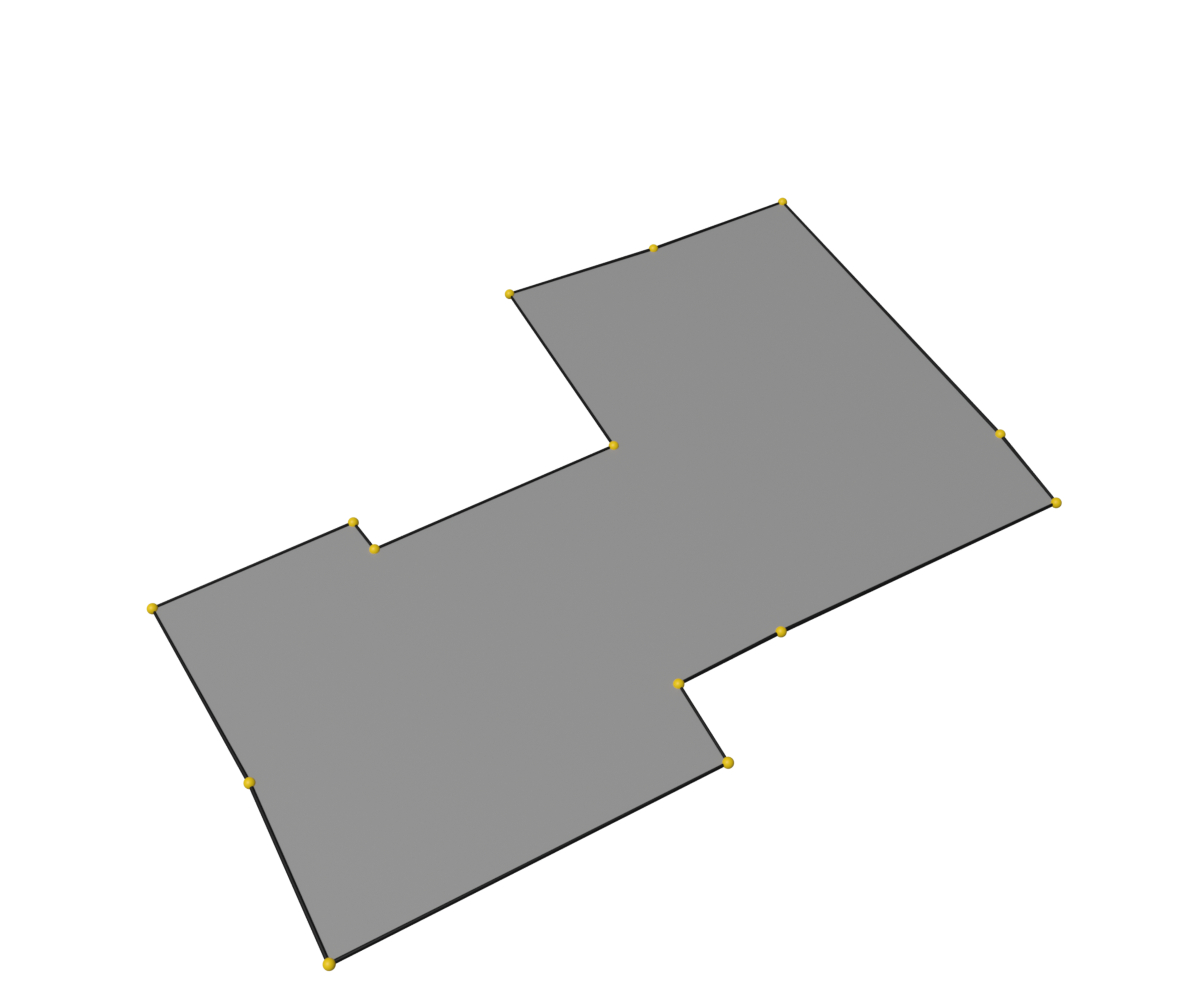}}
        \end{overpic}
    \caption{Input}
    \label{subfig:pipeline_a}
    \end{subfigure}
    &

    \begin{subfigure}{\mywidth}
        \begin{tikzpicture}[spy using outlines={every spy on node/.append style={smallwindow}}]
    
            \coordinate (ai) at (1.05,0.0);
            \coordinate (aj) at (1.2,0.5);

            \coordinate (bi) at (-1.0,-0.15);
            \coordinate (bj) at (-1.0,3.0);


          \node[anchor=south] (FigA) at (0,0){%
    \includegraphics[width=\linewidth,mytrim]{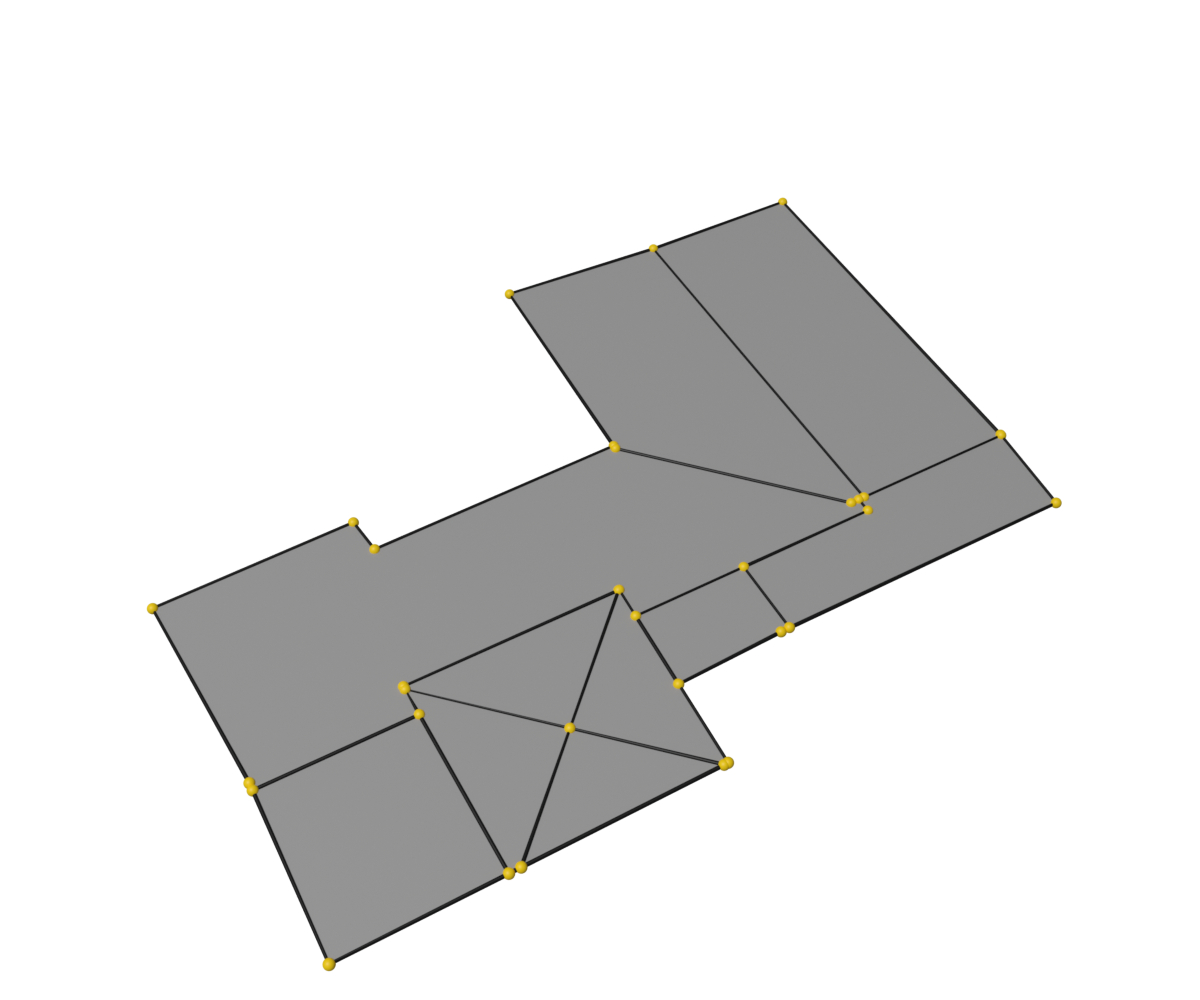}};
          \spy [closeup,magnification=3] on ($(FigA)+(ai)$) 
              in node[largewindow] at ($(FigA.south)+(aj)$);
          \spy [closeup,magnification=3] on ($(FigA)+(bi)$) 
              in node[largewindow] at ($(FigA.south)+(bj)$);
        \end{tikzpicture}
        \caption{2D polygonal partition}
        \label{subfig:pipeline_b}
    \end{subfigure}

    &

    \begin{subfigure}{\mywidth}
        \begin{tikzpicture}[spy using outlines={every spy on node/.append style={smallwindow}}]
    
          \node[anchor=south] (FigA) at (0,0){%
          \includegraphics[width=\linewidth,mytrim]{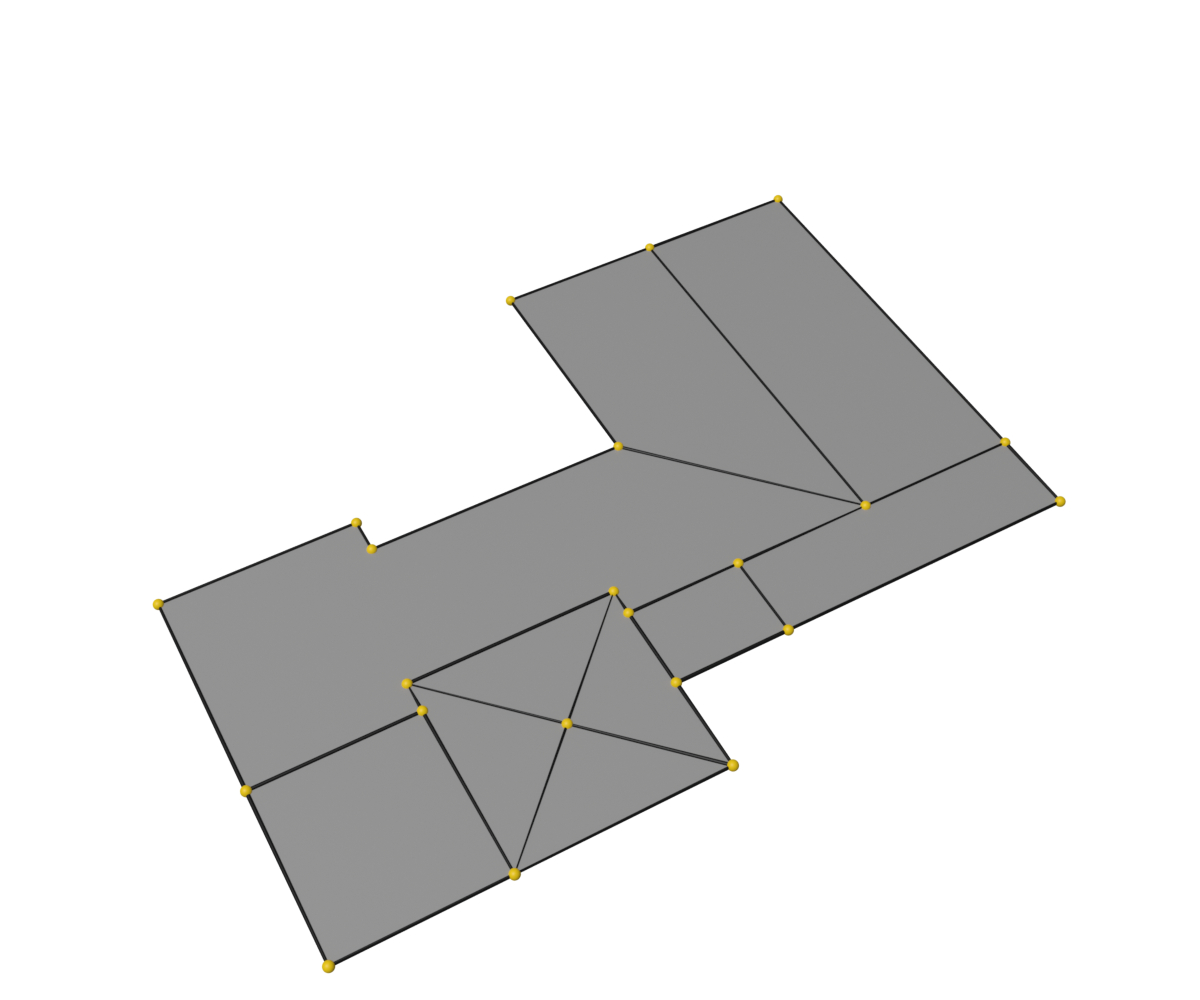}};
          \spy [closeup,magnification=3] on ($(FigA)+(ai)$) 
              in node[largewindow] at ($(FigA.south)+(aj)$);
          \spy [closeup,magnification=3] on ($(FigA)+(bi)$) 
              in node[largewindow] at ($(FigA.south)+(bj)$);
        \end{tikzpicture}
        \caption{Regularized partition}
        \label{subfig:pipeline_c}
    \end{subfigure}

    &
    \begin{subfigure}{\mywidth}
    \includegraphics[width=\linewidth,mytrim]{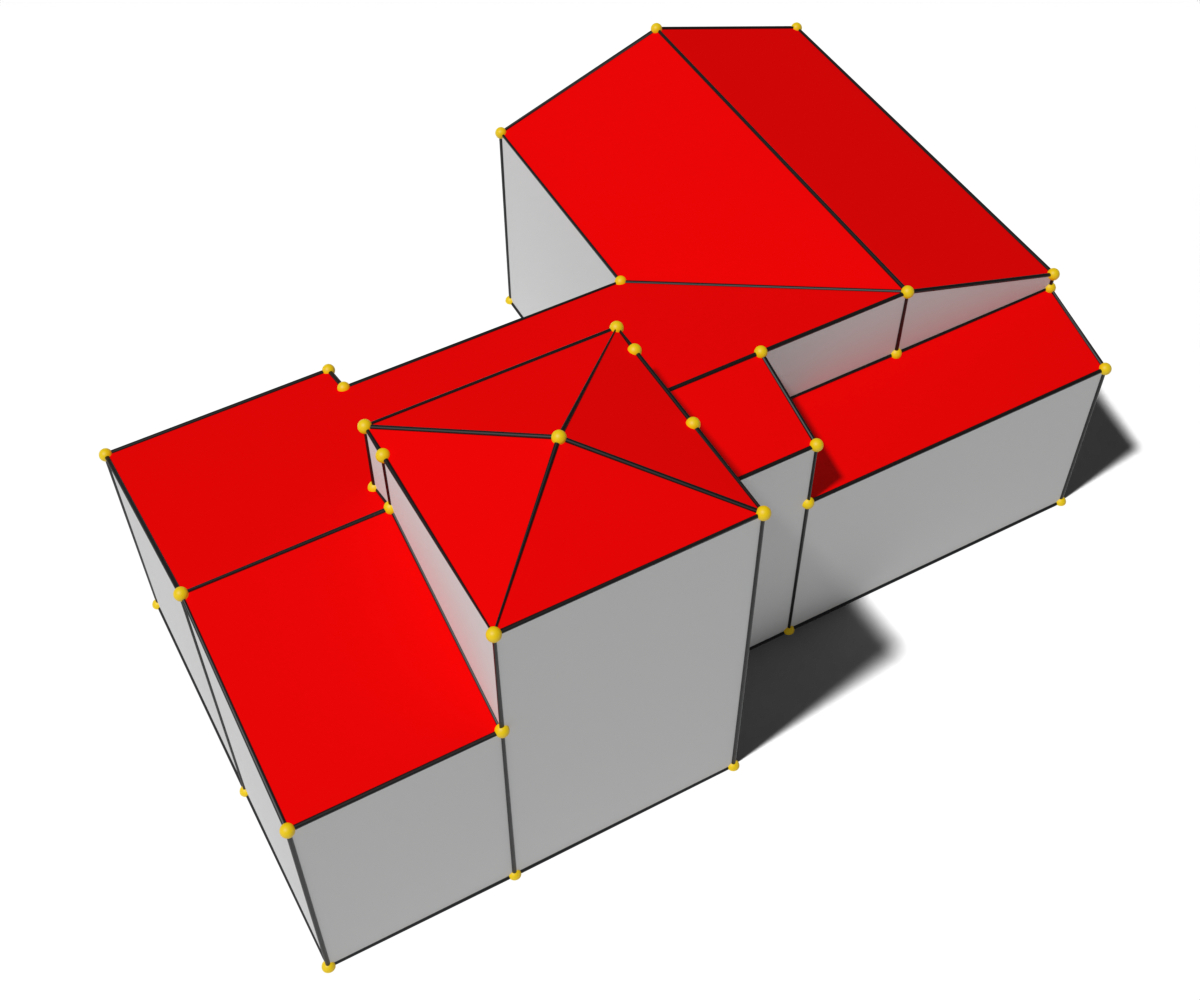}
    \caption{Regularized extrusion}
    \label{subfig:pipeline_d}
    \end{subfigure}
    
    \end{tabular}
    \vspace{-0.2cm}
    \caption{\textbf{Overview.} Starting from a LiDAR scan and a building footprint \subref{subfig:pipeline_a}, we first construct a 2D polygonal partition of the roof structure
    (\subref{subfig:pipeline_b}, Section \ref{sec-step1}, Fig.~\ref{fig:step1}). The partition is then regularized to both enforce orthogonality, parallelism and collinearity between edges and simplify the vertex layout (\subref{subfig:pipeline_c}, Section \ref{sec-step2}, Fig.~\ref{fig:step2}). Finally, the partition is extruded using an optimization procedure that preserves the planarity of roof sections and the horizontality of rooftop edges
    (\subref{subfig:pipeline_d}, Section \ref{sec-step3}, Fig.~\ref{fig:step3}).}
    \label{fig:overview}
    \end{figure*}
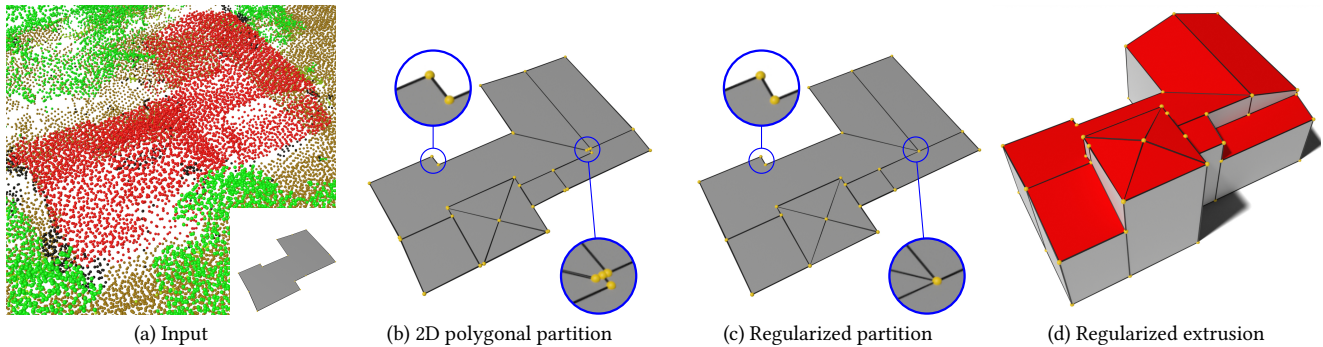
    
\vspace{-0.2cm}
\paragraph{Plane assembly.} Detecting planes from the input data and connecting them into a mesh is also a popular strategy. 

Plane detection is traditionally performed using Hough transform \cite{overby2004automatic}, RANSAC \cite{schnabel2007efficient} or region-growing procedures \cite{rabbani2006segmentation,lafarge2012creating}, all these methods requiring the tuning of a few parameters. 
Neural architectures \cite{zhang2022improved, jiang2020pointgroup, li2022point2roof, yan2021hpnet}, which are parameter-free, show promising results. Trained on synthetic CAD databases, they however do not generalize well on real-world data. Yu \etal \cite{yu2022finding} propose an energy-based model in which the retrieved planes are iteratively refined to satisfy coverage or compactness conditions.



Subsequent plane assembling techniques then either construct a connectivity graph between the detected planes \cite{bouzas2020structure,chen2008architectural,yang2022connectivity} to find out the vertices and edges composing the output mesh, or compute a space decomposition by plane slicing operations before extracting a mesh from it \cite{chauve2010robust, nan2017polyfit, bauchet2020kinetic, jiang2023structure}. 
In particular, PolyFit \cite{nan2017polyfit} has been used in several building reconstruction pipelines \cite{xie2021combined, huang2022city3d, liu2019topolap}. However, due to the complexity of its binary linear optimization, these approaches do not scale well to complex structures composed of a few dozens or hundred planar components. The space decomposition can also be done in 2D through a planimetric arrangement which is then extruded to 3D \cite{chen2017topologically,poullis20113d,li2019modelling,peters2022geoflow,lafarge2012creating}. These solutions offer a good accuracy but often produce overly-complex output meshes, \eg with tiny facets that adjust the connection between four planes or more.



\vspace{-0.2cm}

\paragraph{Neural models.} Plane assembly based methods can also be combined with a learned occupancy field \cite{chen2022points2poly}. 
Such methods require a minimal number of points for vertical surfaces to deliver accurate results, which is hard to guarantee in practice. Furthermore, the learned occupancy field does not generalize well to unknown complex building types. 
Liu \etal \cite{liu2024point2building} propose a generative model that predicts sequences of vertices and faces that form the output polygonal mesh. However, the output does not adhere to strong geometric guarantees.

\vspace{-0.2cm}
\paragraph*{Building regularization.} Reconstructing 3D models that preserve geometric regularities contained in the building structures can be addressed by plane assembly methods. One strategy consists in regularizing configurations of planes either after their detection \cite{cgal_shape_detection,li2011globfit} or during \cite{monszpart2015rapter,oesau_cgf16, yu2022finding}. Exploited in \cite{zhou_cvpr12,verdie_tog15,rs13010129} with roof symmetry and facade orthogonality and parallelism, this solution helps but does not guarantee highly-regular output meshes. This can also be done in 2D with line-segments \cite{bauchet2018kippi,cgal_shape_regularization} or polygons \cite{huang2022city3d} describing the building contours. To our knowledge, only Vuillamy et al. \cite{vuillamy_cgf22} try to regularize a space partition directly. They simplify 2D polygonal partitions by encouraging line concurrency and orthogonality only. Moreover, their point-line projective duality formulation requires high computing resources and does not scale to complex partitions.

\section{Proposed method}
\label{ourmethod}

\subsection{Overview}

Our algorithm takes as input (i) an airborne Lidar point cloud, and (ii) a set of polygonal footprints describing the contours of buildings. The latter can be either obtained via online cadaster map databases, \eg \cite{openstreetmap}, or computed by an automatic building contouring method \cite{albers2016automatic,bauchet2019city,yang2013automated}.

The output 3D models are polygon surface meshes that are, by construction, watertight, 2-manifold and intersection-free. The polygonal facets can be decomposed into triangles using a constrained Delaunay triangulation with the guarantee that triangles from the same polygonal facet are exactly co-planar. This triangle decomposition is used later in Section \ref{experiments} to fairly compare the mesh complexity with competitors.

\figref{fig:overview} illustrates the three main steps of our algorithm, namely (i) the construction of 2D polygonal partitions describing the roof structures, (ii) the regularization and simplification of these partitions, and (iii) the extrusion of the partitions to 3D.

\begin{figure*}[t]
    \captionsetup[figure]{position=auto}
    \newcommand{\mywidth}{0.23\linewidth}
    \newcommand{\mywidthd}{0.38\linewidth}
    \definetrim{mytrim}{70 0 70 0}
    \definetrim{mytrimd}{20 70 40 70}
    
    \begin{tabular}{@{}cccc@{}}
    \centering

    \begin{subfigure}{\mywidth}
    \includegraphics[width=\linewidth,mytrim]{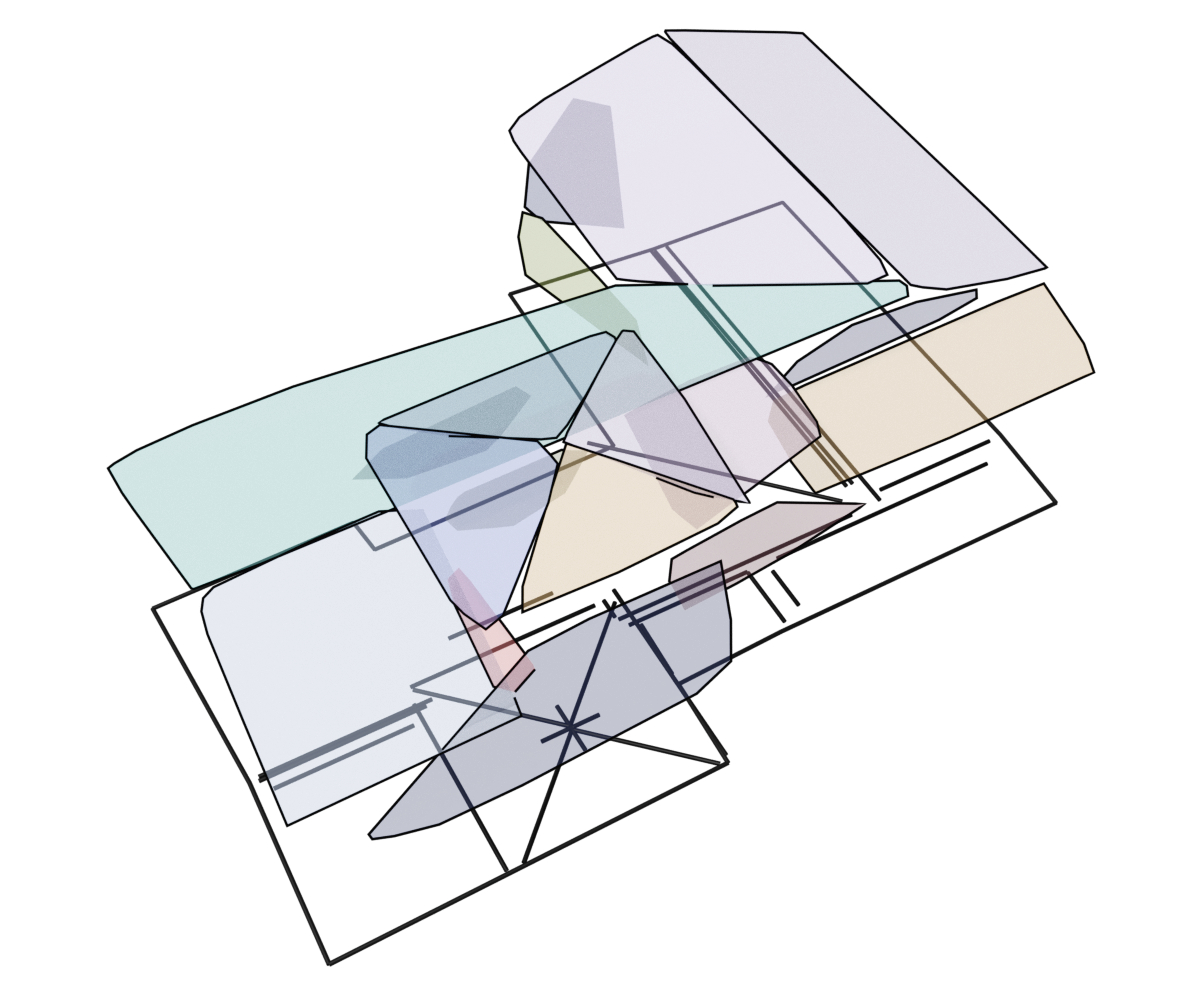}
    \caption{Line-segment projection}
    \label{subfig:step1_a}
    \end{subfigure}
    &
    \begin{subfigure}{\mywidth}
    \includegraphics[width=\linewidth,mytrim]{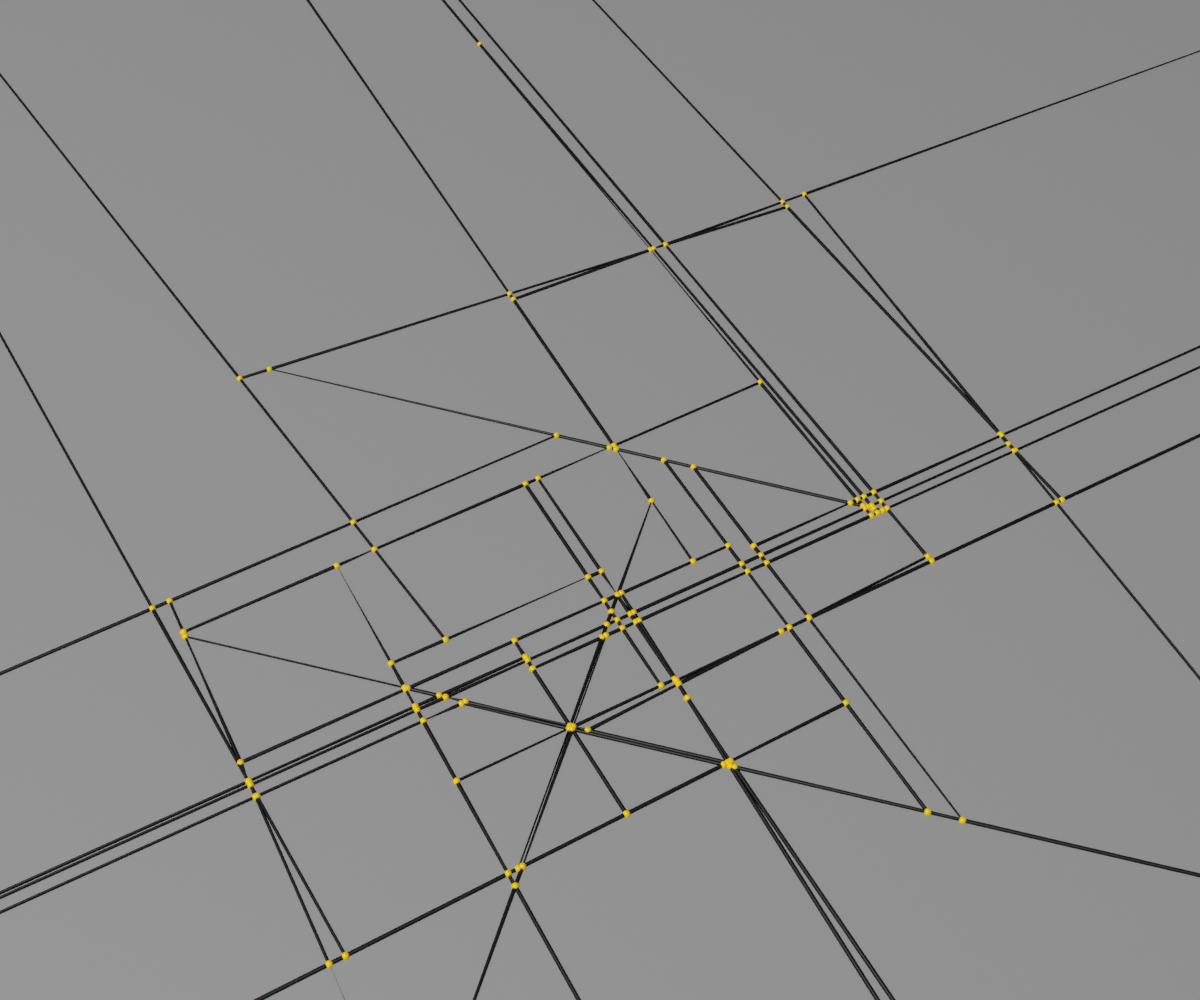}
    \caption{Initial partition}
    \label{subfig:step1_b}
    \end{subfigure}
    &
    \begin{subfigure}{\mywidth}
    \includegraphics[width=\linewidth,mytrim]{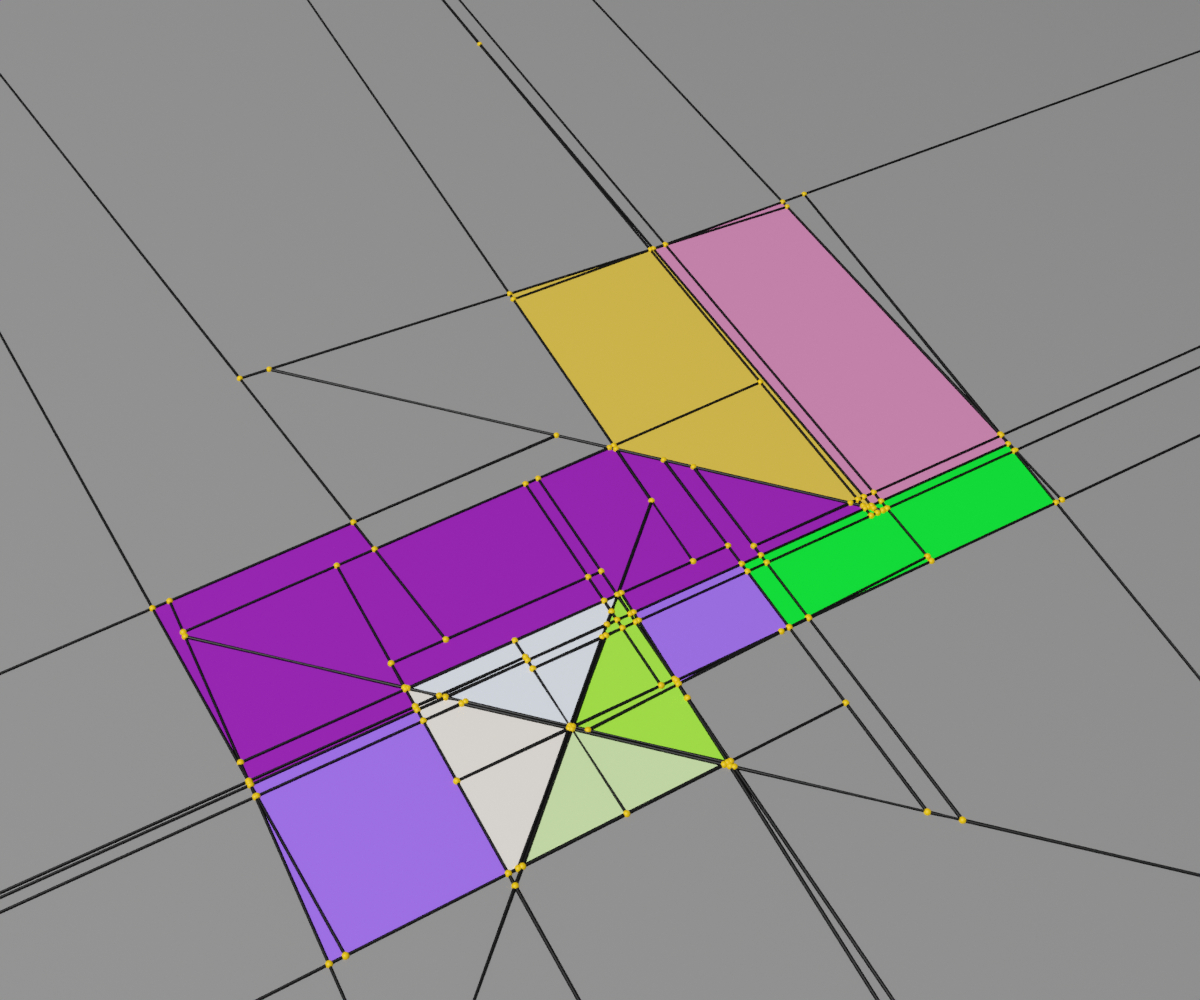}
    \caption{Labeling}
    \label{subfig:step1_c}
    \end{subfigure}
    &
    \begin{subfigure}{\mywidth}
    \includegraphics[width=\linewidth,mytrim]{figures/pipeline/Node_69805_footprint_partitioned.jpg}
    \caption{2D polygonal partition}
    \label{subfig:step1_d}
    \end{subfigure}
    
    \end{tabular}
    \vspace{-0.2cm}
    \caption{\textbf{Construction of 2D polygonal partition.} Line-segments at the intersection of adjacent detected planes and on their boundary are projected into the horizontal plane, jointly with the input footprint line-segments \subref{subfig:step1_a}. An initial, dense 2D  polygonal partition is built from all these line-segments by kinetic simulation \subref{subfig:step1_b} and enriched by plane labels (\subref{subfig:step1_c}, colored polygons). Polygonal cells with same label are then regrouped to form a 2D polygonal partition that describes the roof structure \subref{subfig:step1_d}.}
    \label{fig:step1}
    \end{figure*}

\subsection{Construction of 2D polygonal partitions} \label{sec-step1}
The first step consists in constructing a 2D polygonal partition that represents a projection of the roof structure to the horizontal plane. This partition is not purely geometric, but is enriched by 3D information. Each polygonal cell of the partition is associated with a 3D plane approximating the corresponding roof section in 3D space.
We start by detecting 3D planes from the input point cloud. The detection is controlled by two parameters: a fitting tolerance that specifies the maximal distance of an inlier point to its associated plane, and a minimal number of inliers per plane that avoids the detection of too small components. 
Because airborne laser scans are usually acquired at a near-nadir angle they include only little information on vertical components of the building. Therefore, the detected planes mostly correspond to non-vertical roof sections of the building. To also recover vertical sections we continue as follows.
For each detected plane, we compute a 3D planar alpha shape \cite{edelsbrunner1983shape} as the 2D alpha shape of a set of inlier points projected into the plane, later referred to as a planar primitive.
We now extract two types of 3D line segments from the configuration of planar primitives. The first type corresponds to a potential connection between two adjacent roof sections. These line segments, referred as intersection lines, are computed as the intersections between the pairs of adjacent planar primitives. The second type, referred as discontinuity lines, corresponds to vertical discontinuities in the roof structure. These line segments are a subset of edges composing the planar primitive contours after simplification \cite{neyer1999line}. We use all edges whose difference of the average heights of the points located on both sides is greater than a given threshold (set to 50~cm in our experiments). 
The sets of intersection and discontinuity lines are then projected to the horizontal plane together with the lines forming the input polygonal footprint (\figref{subfig:step1_a}). We then extend all the line segments at constant speed within a kinetic simulation \cite{bauchet2018kippi} to form an initial 2D polygonal partition (\figref{subfig:step1_b}). 

Next, we assign a label to each cell of the partition that corresponds to one of the detected planes or to the ground. This assignment procedure is formulated as an energy minimization problem with discrete variables. Let $C = (c_1, c_2 \hdots c_N)$ be a polygonal partition with $N$ cells, and $L = \{0, 1 \hdots M\}$ be a vector of plane indices, where $M$ is the number of planar primitives initially extracted. Each cell $c_i$ defines a discrete variable $x_i \in L$. $x_i = 0$ means that the related cell is not assigned to any of the planes, which can be the case if $c_i$ is outside the building footprint, for instance. We obtain an optimal label assignment $X^* = (x_1, x_2 \hdots x_N)$ by minimizing the energy
\begin{equation}
E(X) = E_d(X) + E_p(X) + E_c(X)
\end{equation}
where $E_d(X)$ is a data term encouraging a variable $x_i$ to accept a label that corresponds to a planar primitive that is next to the cell $c_i$. $E_p(X)$ is a pairwise smoothness term penalizing a label difference for adjacent cells $c_i$ and $c_j$, based on the height difference on the edge that is common to those cells. Note that this definition does not penalize a difference of labels in ridge and hip lines, where two planes intersect in the 3D space. $E_c(X)$ is a complexity term, set to the total number of edges required, to describe the $M$ polygons resulting from the fusion of all cells with same labels.
Because $E(X)$ is not convex and lives in a discrete domain, its optimization is a difficult problem. We search for an approximate solution using an iterative scheme. Given an initial label assignment $X_0$, we generate a set of neighbor configurations, in which one or several cells from the partition, are transferred from one current roof section to another. Configurations associated with an energy decrease are sorted in a priority queue. Then, we pop from the queue the configuration that corresponds to the largest energy drop, and repeat the process until the queue gets empty.
Each cell of the initial partition with a cell label other then ground now  corresponds to a roof section and is used as a final polygon in the 2D polygonal partition (\figref{subfig:step1_d}).

\subsection{Regularization of 2D polygonal partitions} \label{sec-step2}
 
\begin{figure*}[t]
    \captionsetup[figure]{position=auto}
    \newcommand{\mywidth}{0.25\linewidth}
    \newcommand{\mywidthd}{0.38\linewidth}
    \definetrim{mytrim}{70 0 70 0}
    \definetrim{mytrimd}{20 70 40 70}

    \tikzstyle{closeup} = [
  opacity=1.0,          
  height=1.0cm,         
  width=1.0cm,          
  connect spies, blue  
]
\tikzstyle{largewindow} = [circle, blue, line width=0.3mm]
\tikzstyle{smallwindow} = [circle, blue,line width=0.15mm]

    \begin{tabular}{@{}c@{}c@{}c@{}c@{}}
    \centering

    \begin{subfigure}{\mywidth}
        \begin{tikzpicture}[spy using outlines={every spy on node/.append style={smallwindow}}]
    
            \coordinate (ai) at (1.05,0.0);
          \coordinate (aj) at (-1.2,3.0);
    
          \coordinate (bi) at (-0.35,-1.55);
          \coordinate (bj) at (1.2,0.5);
    
          \node[anchor=south] (FigA) at (0,0){%
          \includegraphics[width=\linewidth,mytrim]{figures/pipeline/Node_69805_footprint_partitioned.jpg}};
          \spy [closeup,magnification=3] on ($(FigA)+(ai)$) 
              in node[largewindow] at ($(FigA.south)+(aj)$);
          \spy [closeup,magnification=3] on ($(FigA)+(bi)$) 
              in node[largewindow] at ($(FigA.south)+(bj)$);
        \end{tikzpicture}
        \caption{2D polygonal partition}
        \label{subfig:step2_a}
    \end{subfigure}

    &

    \begin{subfigure}{\mywidth}
        \begin{tikzpicture}[spy using outlines={every spy on node/.append style={smallwindow}}]  
          \node[anchor=south] (FigA) at (0,0){%
          \includegraphics[width=\linewidth,mytrim]{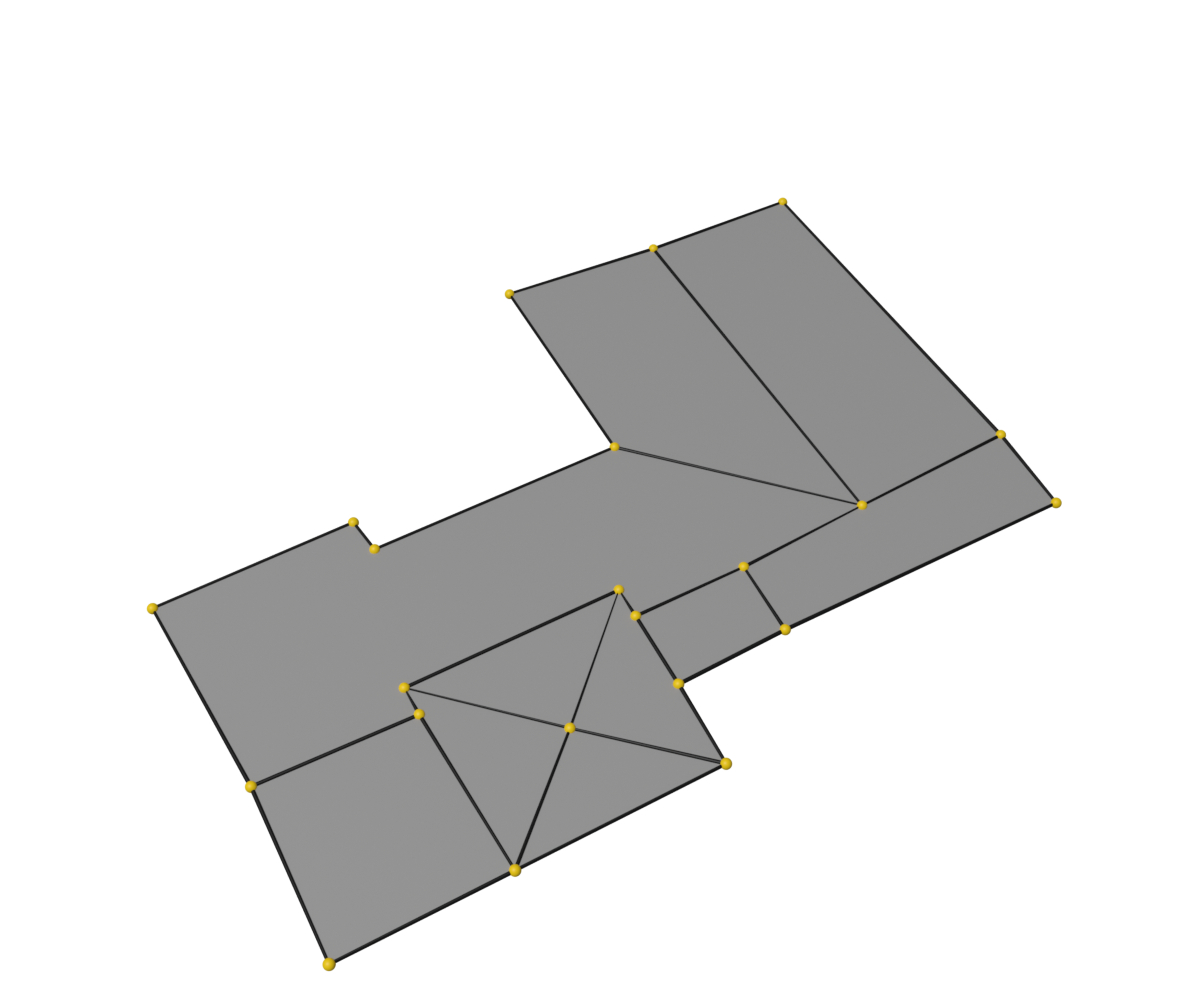}};
          \spy [closeup,magnification=3] on ($(FigA)+(ai)$)     
            in node[largewindow] at ($(FigA.south)+(aj)$);
          \spy [closeup,magnification=3] on ($(FigA)+(bi)$) 
              in node[largewindow] at ($(FigA.south)+(bj)$);
        \end{tikzpicture}
        \caption{Vertex simplification}
        \label{subfig:step2_b}
    \end{subfigure}

    &

    \begin{subfigure}{\mywidth}
        \begin{tikzpicture}[spy using outlines={every spy on node/.append style={smallwindow}}]
    
          \coordinate (ai) at (-1.0,-0.15);
          \coordinate (aj) at (-1.0,3.0);
    
          \coordinate (bi) at (-1.4,-1.15);
          \coordinate (bj) at (1.2,0.5);
    
          \node[anchor=south] (FigA) at (0,0){%
          \includegraphics[width=\linewidth,mytrim]{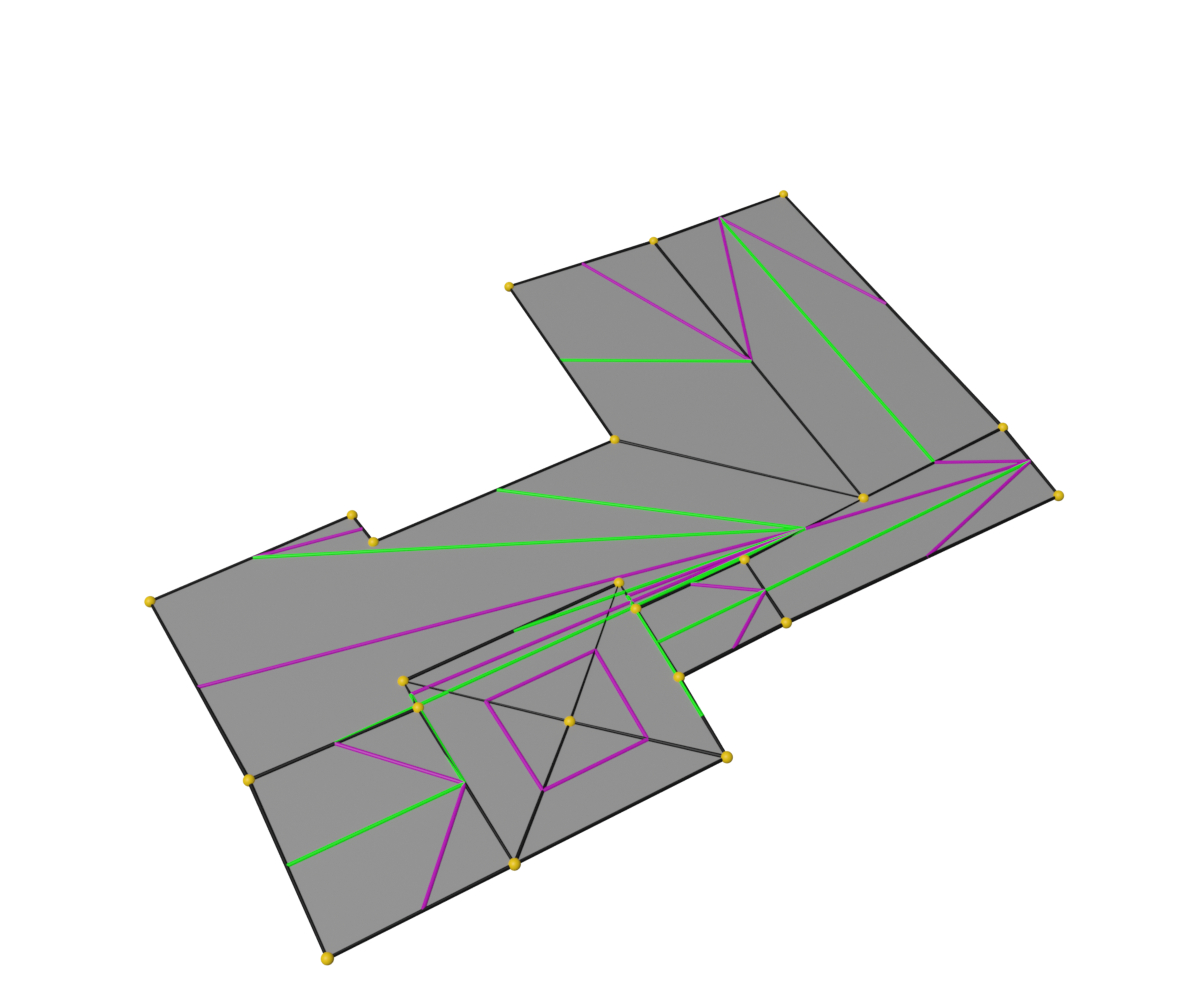}};
          \spy [closeup,magnification=3] on ($(FigA)+(ai)$) 
              in node[largewindow] at ($(FigA.south)+(aj)$);
          \spy [closeup,magnification=3] on ($(FigA)+(bi)$) 
              in node[largewindow] at ($(FigA.south)+(bj)$);
        \end{tikzpicture}
        \caption{Regularity detection}
        \label{subfig:step2_c}
    \end{subfigure}

    &

    \begin{subfigure}{\mywidth}
        \begin{tikzpicture}[spy using outlines={every spy on node/.append style={smallwindow}}]
    
          \node[anchor=south] (FigA) at (0,0){%
          \includegraphics[width=\linewidth,mytrim]{figures/pipeline/Node_69805_footprint_optimised.jpg}};
          \spy [closeup,magnification=3] on ($(FigA)+(ai)$) 
              in node[largewindow] at ($(FigA.south)+(aj)$);
          \spy [closeup,magnification=3] on ($(FigA)+(bi)$) 
              in node[largewindow] at ($(FigA.south)+(bj)$);
        \end{tikzpicture}
        \caption{Optimization}
        \label{subfig:step2_d}
    \end{subfigure}

    
    \end{tabular}
    \vspace{-0.2cm}
    \caption{\textbf{Regularization of 2D polygonal partition.} The 2D polygonal partition \subref{subfig:step2_a} is first simplified by collapsing small edges (\subref{subfig:step2_b}, see close-ups). The regularity graph $G$ is then built by detecting pairs of near-parallel edges (green lines) and near-orthogonal edges (purple lines) in \subref{subfig:step2_c}. A global optimization of the vertex coordinates constrained by $G$ is then performed to regularize the partition (\subref{subfig:step2_d}).}
    \label{fig:step2}
    \end{figure*}

The second step of our algorithm aims to simplify the 2D polygonal partition and enhance its regularity. We address this problem with a global optimization procedure under geometric constraints.
 
Many building roof types include corner types that require four or more planar components to meet in the same point. However, 3D planes detected from Lidar point clouds do not exhibit such a behaviour. \figref{subfig:step2_a} shows how this problem creates an overly complex polygonal partition, leading to the presence of extra vertices and small facets in the final output mesh.  
To solve this issue, we start by collapsing short edges with a length smaller $\tau_h$ in the 2D polygonal partition (\figref{subfig:step2_b}). To enhance the regularity of the partition we detect near parallel and orthogonal edges. We then build a graph $G$, where each node represents an edge of the partition, and an edge in $G$ represents a near-parallel or near-orthogonal pair of edges in the partition (\figref{subfig:step2_c}). In practice, we compute $G$ before collapsing short edges, as the edge collapse can significantly alter edge orientations.
We finally apply a global optimization procedure to find new positions for all vertices of the partition that respect the parallel and orthogonal constraints of adjacent edges (\figref{subfig:step2_d}). Note that the 3D plane equation associated with each cell of the initial partition is preserved by these geometric operations. 

\subsection{Extrusion}  \label{sec-step3}

\begin{figure*}[t]

    \tikzstyle{closeup} = [
  opacity=1.0,          
  height=1.0cm,         
  width=1.0cm,          
  connect spies, blue  
]
\tikzstyle{largewindow} = [circle, blue, line width=0.3mm]
\tikzstyle{smallwindow} = [circle, blue,line width=0.15mm]

    \captionsetup[figure]{position=auto}
    \newcommand{\mywidth}{0.25\linewidth}
    \newcommand{\mywidthd}{0.38\linewidth}
    \definetrim{mytrim}{70 0 70 0}
    \definetrim{mytrimd}{20 70 40 70}
    
    \begin{tabular}{@{}c@{}c@{}c@{}c@{}}
    \centering

    \begin{subfigure}{\mywidth}
    \includegraphics[width=\linewidth,mytrim]{figures/pipeline/Node_69805_footprint_optimised.jpg}
    \caption{Regularized partition}
    \label{subfig:step3_a}
    \end{subfigure}
    &
    \begin{subfigure}{\mywidth}
    \begin{tikzpicture}[spy using outlines={every spy on node/.append style={smallwindow}}]

      \coordinate (ai) at (-1.75,-0.35);
      \coordinate (aj) at (-0.6,0.3);

      \coordinate (bi) at (-0.18,0.28);
      \coordinate (bj) at (-1.2,3.8);

      \coordinate (ci) at (1.28,0.9);
      \coordinate (cj) at (1.4,1.3);
      
      \node[anchor=south] (FigA) at (0,0){%
      \includegraphics[width=\linewidth,mytrim]{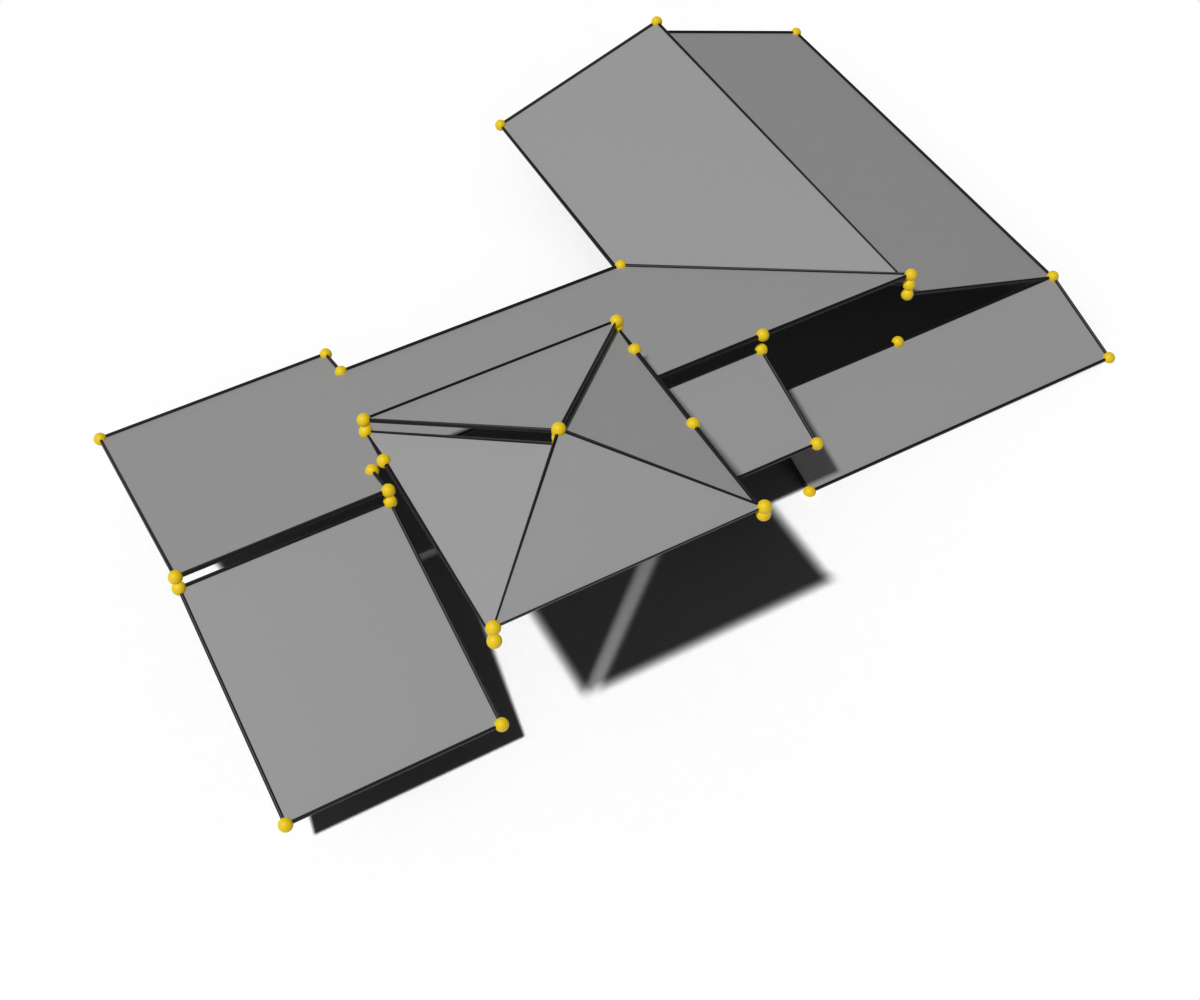}};
      \spy [closeup,magnification=3] on ($(FigA)+(bi)$) 
          in node[largewindow] at ($(FigA.south)+(bj)$);
      \spy [closeup,magnification=3] on ($(FigA)+(ci)$) 
          in node[largewindow] at ($(FigA.south)+(cj)$);
    \end{tikzpicture}
    \caption{Cell extrusion}
    \label{subfig:step3_b}
    \end{subfigure}
    &
    \begin{subfigure}{\mywidth}
      \begin{tikzpicture}[spy using outlines={every spy on node/.append style={smallwindow}}]
        
        \coordinate (ai) at (-1.25,0.12);
        \coordinate (aj) at (-0.1,0.3);

        \node[anchor=south] (FigA) at (0,0){%
        \includegraphics[width=\linewidth,mytrim]{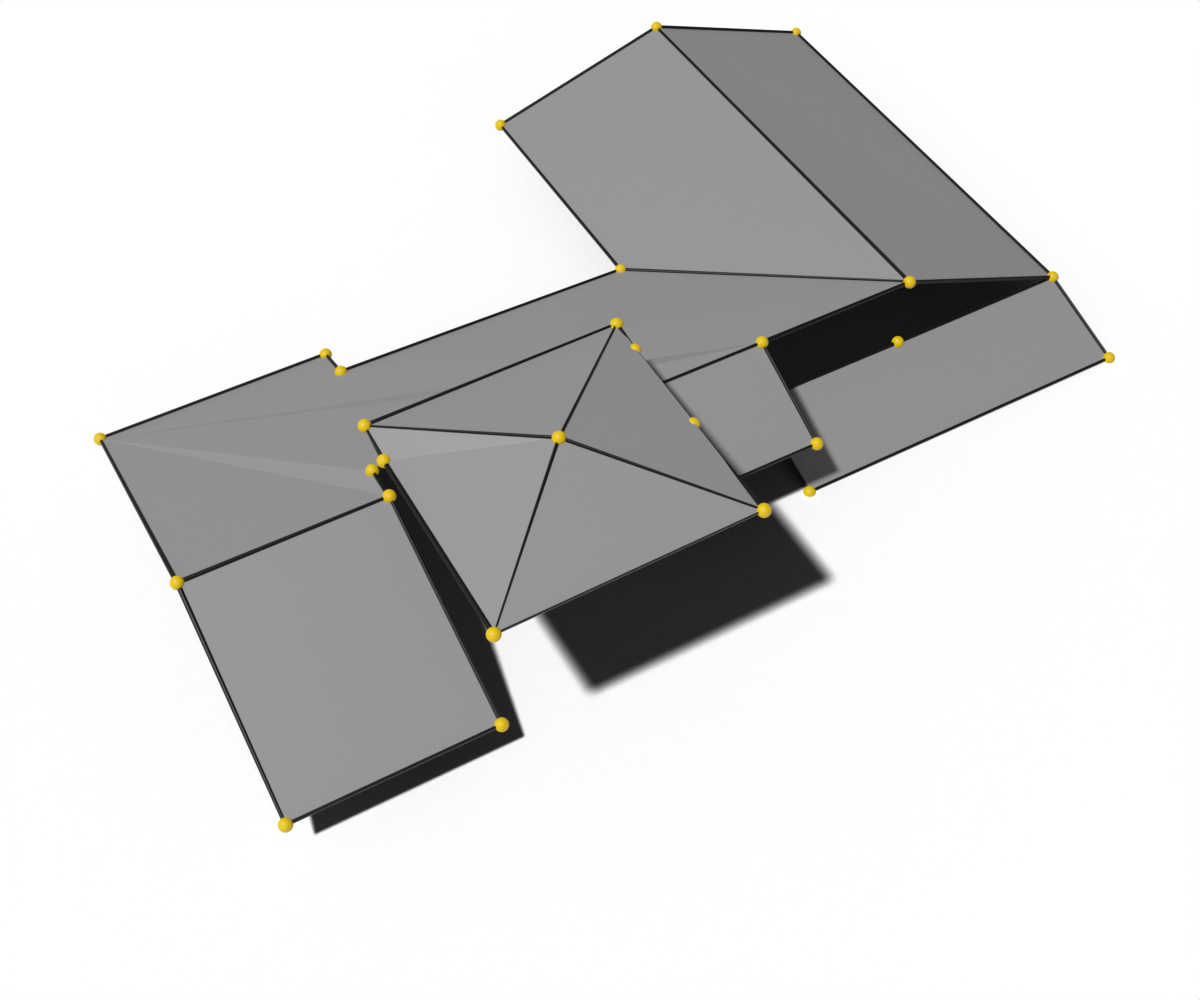}};
        \spy [closeup,magnification=3] on ($(FigA)+(ai)$) 
            in node[largewindow] at ($(FigA.south)+(aj)$);
      \end{tikzpicture}
      \caption{Vertex simplification}
      \label{subfig:step3_c}
      \end{subfigure}
    &
    \begin{subfigure}{\mywidth}
      \begin{tikzpicture}[spy using outlines={every spy on node/.append style={smallwindow}}]
  
        \node[anchor=south] (FigA) at (0,0){%
        \includegraphics[width=\linewidth,mytrim]{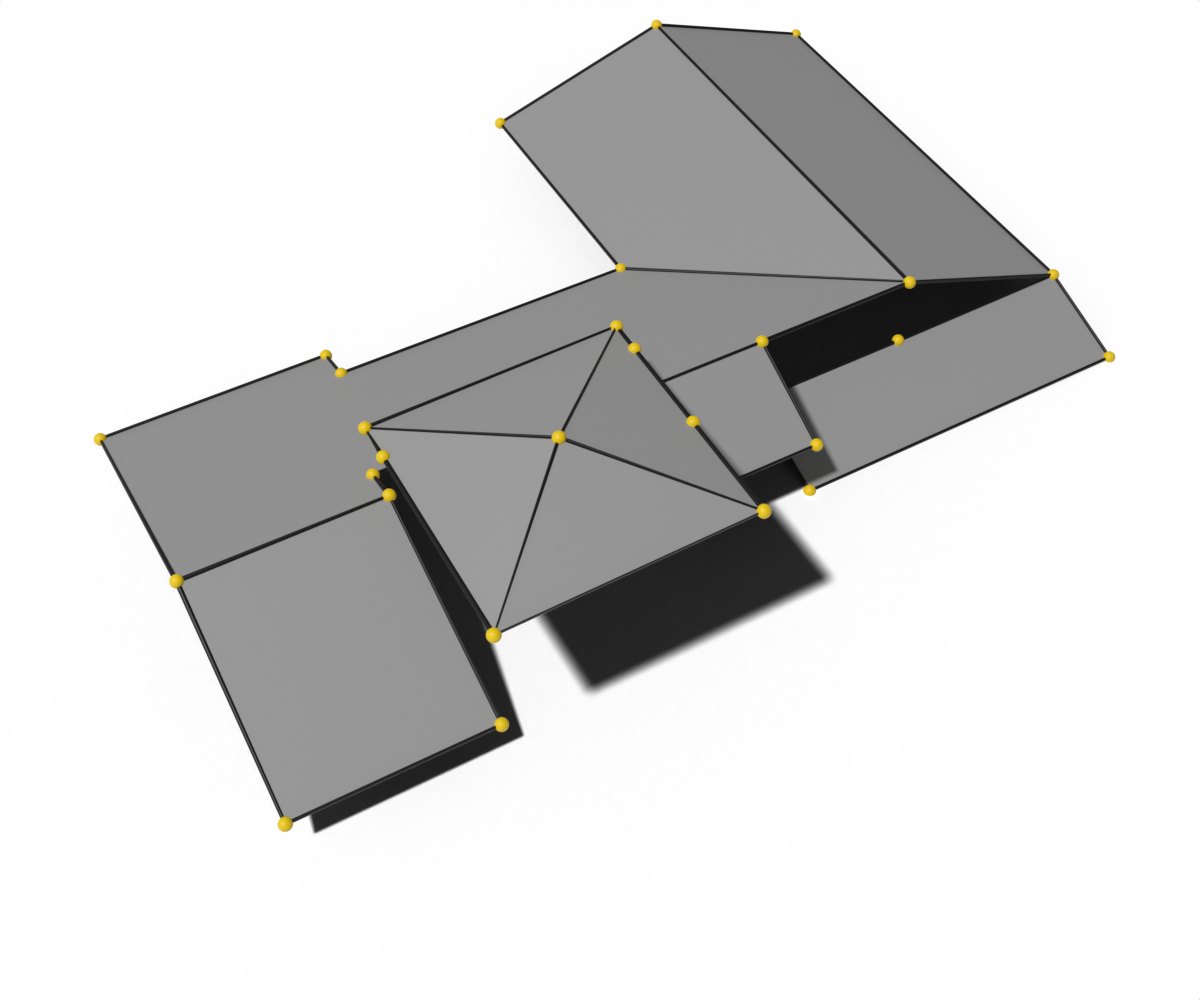}};
        \spy [closeup,magnification=3] on ($(FigA)+(ai)$) 
            in node[largewindow] at ($(FigA.south)+(aj)$);
        \spy [closeup,magnification=3] on ($(FigA)+(bi)$) 
            in node[largewindow] at ($(FigA.south)+(bj)$);
        \spy [closeup,magnification=3] on ($(FigA)+(ci)$) 
            in node[largewindow] at ($(FigA.south)+(cj)$);
      \end{tikzpicture}
      \caption{Optimization}
      \label{subfig:step3_d}
      \end{subfigure}    
    \end{tabular}
    \vspace{-0.2cm}
    \caption{\textbf{Extrusion.} Each cell of the 2D regularized partition \subref{subfig:step3_a} is first extruded according to its associated detected plane \subref{subfig:step3_b}. Close 3D vertices with same x- and y-coordinates are then merged, breaking the planarity of the extruded facets \subref{subfig:step3_c}. The z-coordinates of the vertices are finally readjusted using a global optimization under facet planarity constraints \subref{subfig:step3_d}.}
    \label{fig:step3}
    \end{figure*}

Once regularized, the 2D polygonal partition is extruded by projecting each cell to 3D using the associated plane equations. 
One problem of such an extrusion strategy is that two adjacent roof sections, \eg the two sides of a gabled roof, are not modeled with a continuous transition (Fig.~\ref{subfig:step3_b}). This is because the same vertex in the 2D partition can be extruded to slightly different height values from the different plane equations associated with its incident cells. Simply extruding each vertex to the mean height value of all incident cells breaks the planarity of the reconstructed roof sections (Fig.~\ref{subfig:step3_c}).
To force transition continuity between close adjacent roof sections without breaking their planarity, we propose a global optimization procedure under geometric constraints, similar to the partition regularization problem in Section \ref{sec-step2}.
We first collapse 3D vertices with a height distance smaller $\tau_v$ and then optimize the height value of each vertex with the constraint that all vertices per roof section need to adhere to the same plane equation  (Fig.~\ref{subfig:step3_d}).
Note that, we keep the x- and y-coordinates of the vertices fixed to not break the parallel and orthogonal symmetries enforced during the horizontal optimization. 
After horizontal and vertical optimization we finally assemble the obtained roof polygonal facets and vertical walls to form the 2.5D building model.



\section{Experiments}
\label{experiments}

\subsection{Experimental setup}

\paragraph*{Dataset.}
We evaluate our method on three different datasets:
\begin{itemize}
\item The full test set of the \textit{Building3D Tallinn dataset} \cite{wang2023building3d}, comprising 3472 building models, at an approximate density of 27 points per squared meter;
\item The \textit{Zurich dataset} \cite{swisstopo} with Lidar point clouds sampled at approximately 12 points per squared meter, for a total area of \SI{4}{\kilo\metre\squared}, for which 3177 reference building models \cite{cityofzurich} are provided in a CityGML LOD2.3 format;
\item The \textit{Helsinki dataset} \cite{cityofhelsinki} with Lidar point clouds sampled at a density of 56 points points per squared meter, covering \SI{4}{\kilo\metre\squared} in total, for which 1275 reference building models are provided in a CityGML LOD2.2 format.
\end{itemize}

\paragraph*{Metrics.}
We measure the quality of building models using the following metrics. For complexity, we count the number of vertices $|V|$ and facets $|F|$ of the model, as well as the ratio $E_{<0.5~\text{m}}$ of edges lower than $0.5$ meter. For accuracy, we sample 100k points on the reconstructed models as well as the reference models and measure (i) the one-sided Chamfer distance $\precision$ from reconstruction to reference samples, and (ii) the one-sided Chamfer distance $\recall$ from the input point cloud to the reconstruction samples.
We do not measure the distance from reference to reconstructed models because we find that some of the reference models have interior facets that produce irrelevant distance measures.

%

\paragraph*{Baselines.}
We compare our method with four different baselines. Most of our competitors rely on the extraction of planar shapes from the input point clouds. Though more sophisticated techniques could be used, we extract planes from point clouds using a standard region-growing procedure \cite{lafarge2012creating}, so that the same configuration of planes can be input to our method and all baselines.
\begin{itemize}
    \item \textbf{2.5D Dual Contouring (2.5DC)}~\cite{zhou2010dualcontouring} detects sharp feature points from the input point cloud and triangulates these points to form 2.5D semi-dense mesh models. Vertices of the mesh are then snapped to detected principal directions for regularization. It takes a building point cloud augmented with normals as input.
    \item \textbf{Kinetic Shape Reconstruction (KSR)}~\cite{bauchet2020kinetic} uses planes detected on the input point cloud and computes a polyhedral decomposition. The cells of the decomposition are labelled as inside or outside using point normal orientation. Finally, a building model is extracted as a concise polygon mesh. For a fair comparison, we input the same plane configuration to KSR and our method, including planes for all vertical discontinuities and per line segment of the footprint polygon.
    \item \textbf{Geoflow}~\cite{peters2022geoflow} is a 2.5D extrusion method that works similar to ours, but without any optimization steps to improve the simplicity of building models. It takes a building point cloud and footprint as input.
    \item \textbf{City3D}~\cite{huang2022city3d} is another plane arrangement method based on PolyFit~\cite{nan2017polyfit}. The pipeline detects planes from the input point cloud and from vertical discontinuities. To target the simplicity of models, the input footprint is regularized before planes are extracted from its line segments.
\end{itemize}

\subsection{Results}
\begin{table}[t]
    \setlength{\tabcolsep}{9pt}
\centering
\caption{\textbf{Quantitative evaluation.} 
For each metric and each dataset, we highlight the \colorbox{blue!25}{best} and \colorbox{blue!10}{second best} scores. $^\dagger$We stop the process after a runtime of 5 min per building.
}
\vspace{-0.2cm}
\resizebox{\linewidth}{!}{%
\begin{tabular}{@{}l|lrrrrrHHr@{}}
\toprule
\multicolumn{2}{c}{} & \multicolumn{3}{c}{\emph{Complexity}} & \multicolumn{2}{c}{\emph{Accuracy}}  & \emph{Symmetry} & \emph{Failure}  & \emph{Perform.} \\
\midrule
\multicolumn{2}{c}{}                                & $|V|$         & $|F|$         &$E_{<0.5~\text{m}}$&  $\recall$ &  $\precision$  &  $S$   &  $|X|$         & Time     \\
\multicolumn{2}{c}{} &   &  &  [\%] & [cm] & [cm]  &&  & [s] \\
\midrule
\multirow{4}{*}{\rotatebox[origin=c]{90}{\hspace{-6mm}\emph{Tallinn}}}
&
\rc\textbf{Reference}~\cite{wang2023building3d}     &  \rc  84.6    & \rc47.5       &\rc1.28   &\rc6.68    & \rc-      &           &  0 & \rc-  \\
&\textbf{2.5DC}~\cite{zhou2010dualcontouring}       &    112        & 187           & \tf 2.86 &  11.9     & 81.9      &           &  0 & \tf 0.13 \\
&\textbf{KSR}                                       & \ts 67.8      & \ts 71.4      &    8.35  & \ts 8.07  &      15.9 &           &     3 & 10.4 \\
&\textbf{Geoflow}~\cite{peters2022geoflow}          &    180        & 172           &    28.0  & \tf 7.18  & 36.4      &           &  0 & \ts 4.03 \\
&\textbf{City3D}$^\dagger$~\cite{huang2022city3d}   &    116        & 199           &    25.0  & 9.39      & \tf  7.24 &           & 291   & 49.1 \\
&\textbf{Ours}                                      & \tf 32.2      & \tf 58.2      & \ts 4.39 &     9.34  &  \ts 13.2 &     0.95  &  584  & 4.19 \\
\midrule
\multirow{4}{*}{\rotatebox[origin=c]{90}{\hspace{-6mm}\emph{Zurich}}}
&
\rc\textbf{Reference}~\cite{cityofzurich}           &\rc 54.6       &\rc 77.2       & \rc 4.48  & \rc58.3   & \rc-      &           & -  & \rc -\\
&\textbf{2.5DC}~\cite{zhou2010dualcontouring}       & 140           & 241           &  \tf 2.57 & \tf 16.4  & 50.6      &           &  6 & \tf 0.052 \\
&\textbf{KSR}~\cite{bauchet2020kinetic}             &  116          & \ts  139      &   8.72    & \ts 22.0  &     33.2  &           & 22    & 6.52 \\
&\textbf{Geoflow}~\cite{peters2022geoflow}          & 227           & 220           &   27.0    &   28.1    & 50.0      &           & 97    &  \ts 1.59\\
&\textbf{City3D}$^\dagger$~\cite{huang2022city3d}   & \ts 112       & 193           &   28.3    &   30.6	& \tf 20.0  &           & 501   & 42.3 \\
&\textbf{Ours}                                      & \tf 58.3      & \tf  109      &  \ts 7.48 &    26.6   & \ts 26.0  & 0.21      &    1  & 3.82 \\
\midrule
\multirow{4}{*}{\rotatebox[origin=c]{90}{\hspace{0mm}\emph{Helsinki}}}
&
\rc\textbf{Reference}~\cite{cityofhelsinki}         & \rc203        & \rc 236       &  \rc 8.97 & \rc 66.2  & \rc-      &           &     - &\rc - \\
&\textbf{2.5DC}~\cite{zhou2010dualcontouring}       & \ts      570  &      1045     &  \tf 2.67 & \tf 22.6  & 69.9      &           &     3 & \tf 1.15 \\
&\textbf{KSR}~\cite{bauchet2020kinetic}             &     619       & \ts   704     &   14.5    &    39.8   & \ts 39.2  &           &     0 & 67.9 \\
&\textbf{Geoflow}~\cite{peters2022geoflow}          & 900           & 883           &   23.1    &     34.2  & 54.7      &           & 63    & \ts 6.12\\
&\textbf{Ours}                                      & \tf  214      &  \tf  393     &  \ts 9.5  &\ts 32.3   &  \tf 37.1 & 0.21      &    1  & 6.81 \\
\bottomrule
\end{tabular}
}
\label{tab:tallinn}
\end{table}
\begin{figure*}[ht]
	\definetrim{mytrim1}{0 0 0 0}
	\definetrim{mytrim2}{100 10 100 10}
	\definetrim{mytrim3}{100 50 100 50}
	\definetrim{mytrim4}{0 0 0 0}
	\newcommand{\mywidth}{0.14\linewidth}
	\newcommand{\mywidthb}{0.18\linewidth}
	\newcommand{\myfontsize}{\scriptsize}

	\setlength{\tabcolsep}{0mm}

    \centering
\begin{tabular}{@{}c@{}ccccccc@{}}

\multirow{3}{*}{\rotatebox{90}{\hspace{6mm}Tallinn}}&
\includegraphics[width=\mywidth,mytrim1]{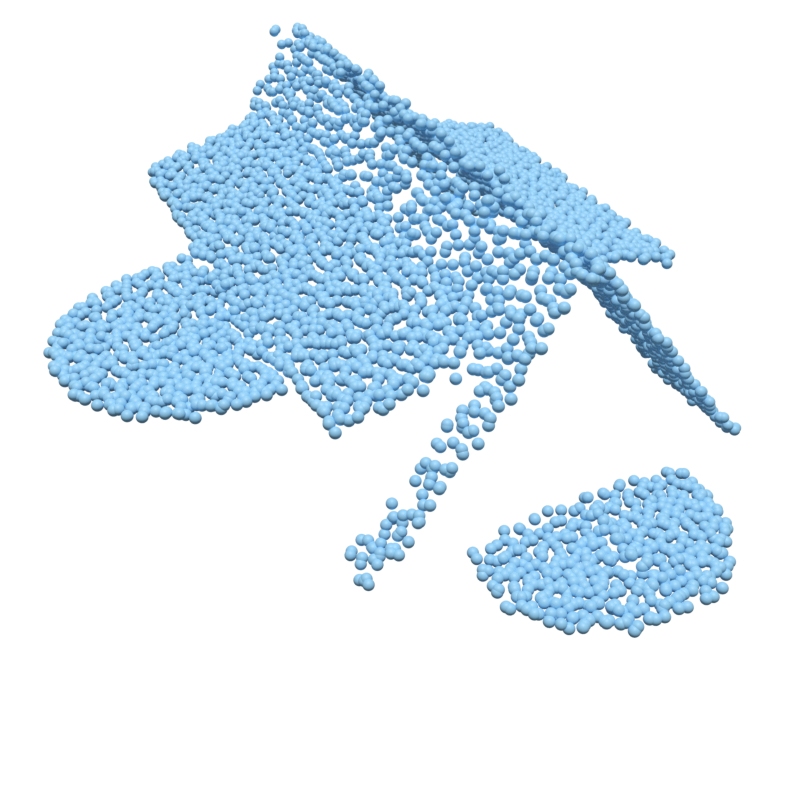}
&
\includegraphics[width=\mywidth,mytrim1]{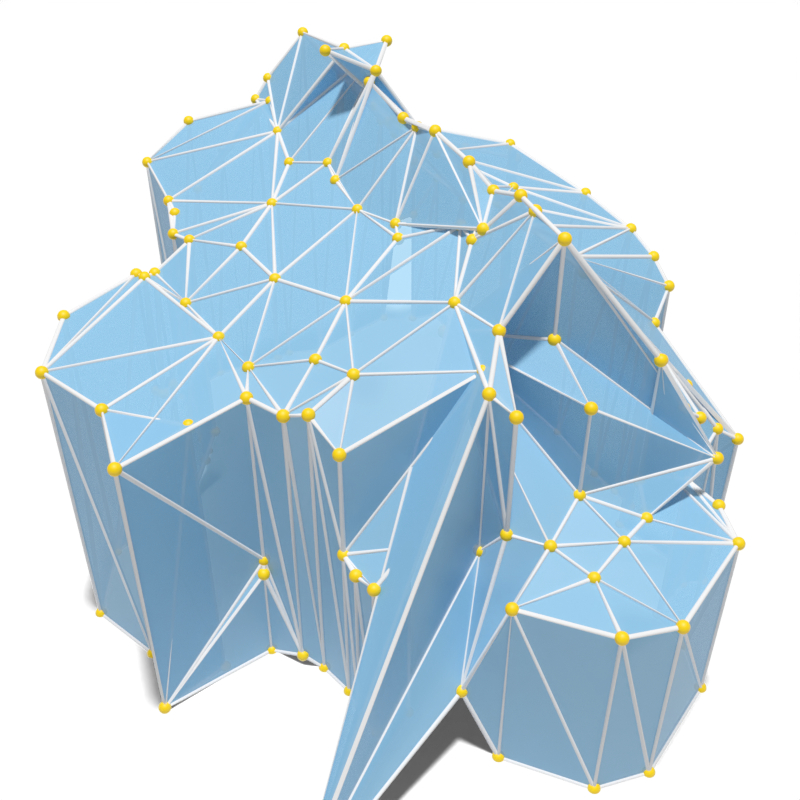}
&
\includegraphics[width=\mywidth,mytrim1]{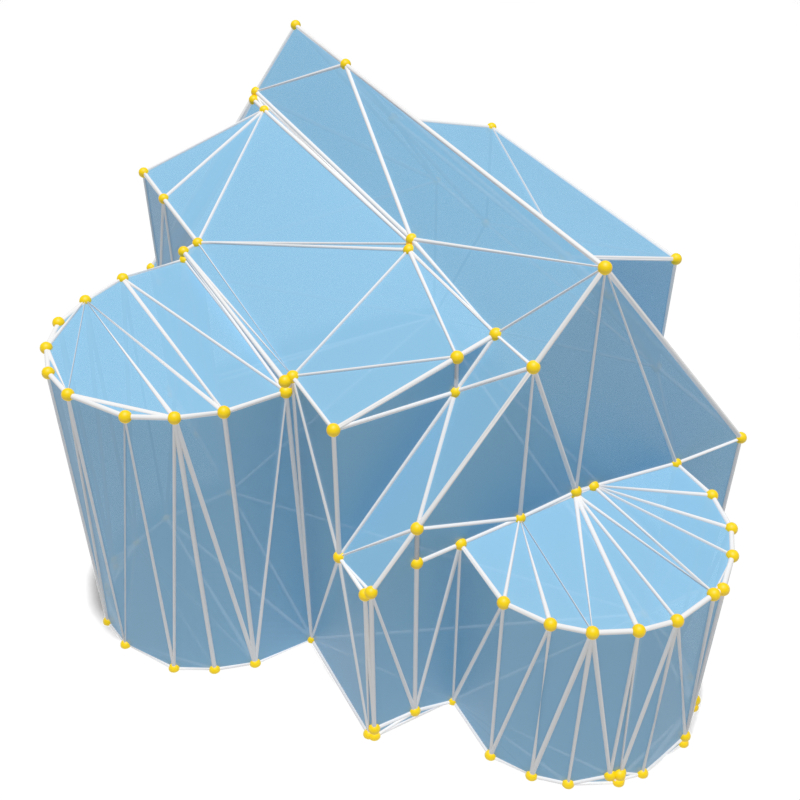}
&
\includegraphics[width=\mywidth,mytrim1]{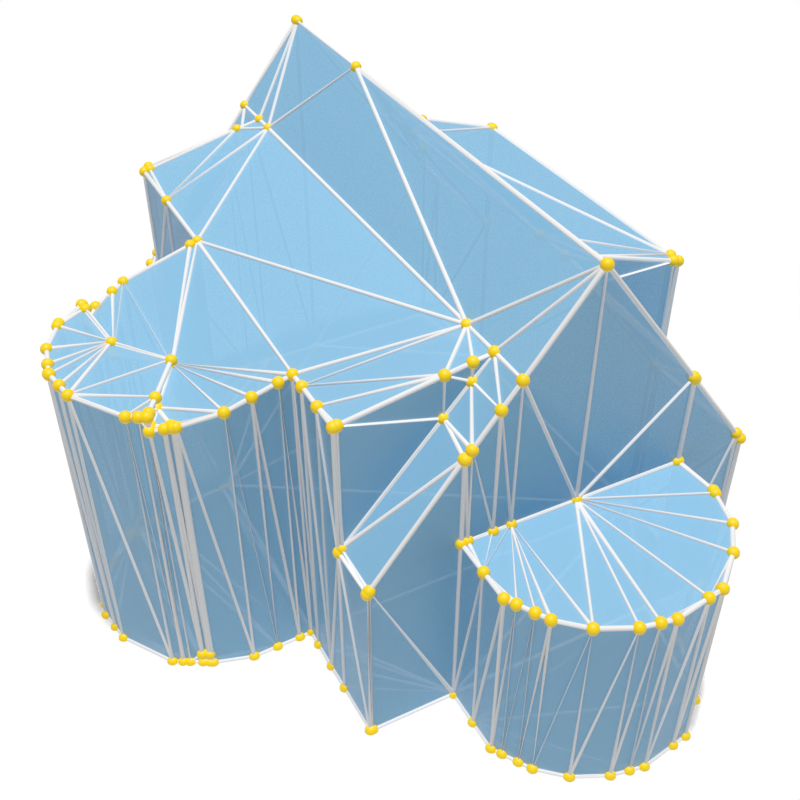}
&
\includegraphics[width=\mywidth,mytrim1]{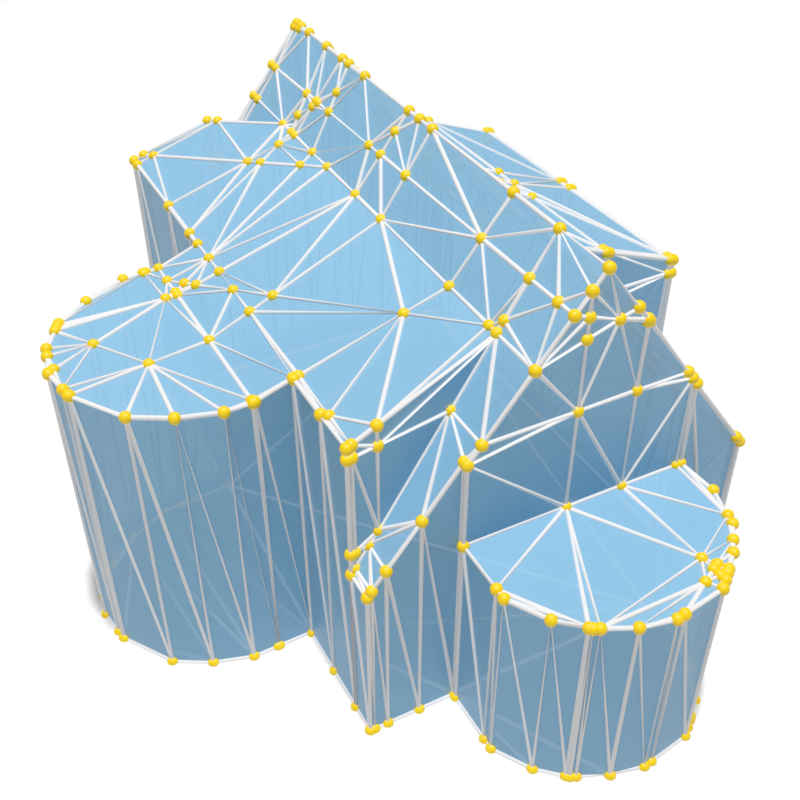}
&
\includegraphics[width=\mywidth,mytrim1]{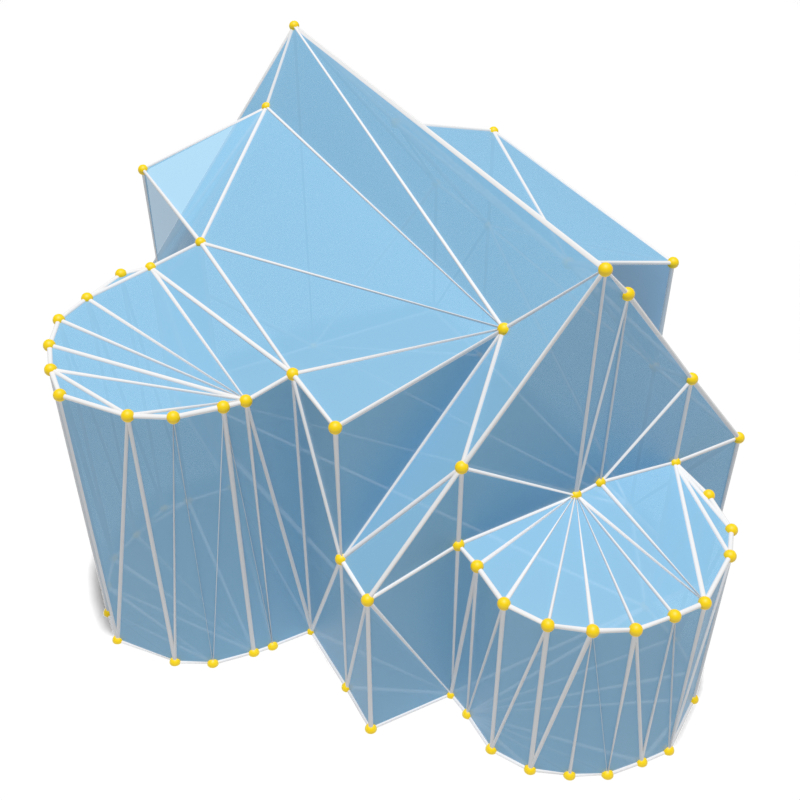}
&
\includegraphics[width=\mywidth,mytrim1]{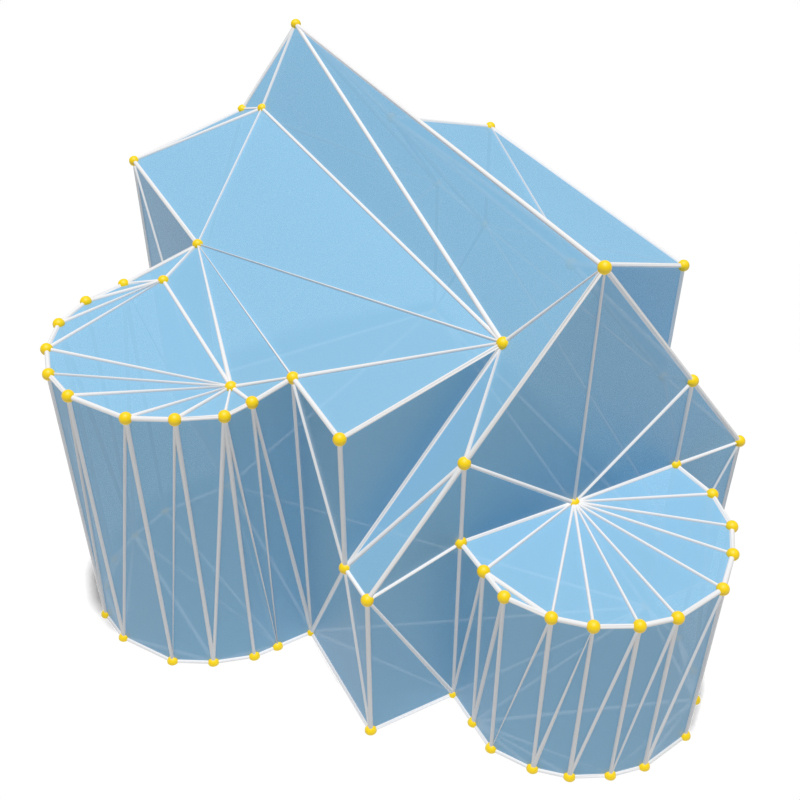}
\\
&
\includegraphics[width=\mywidth,mytrim1]{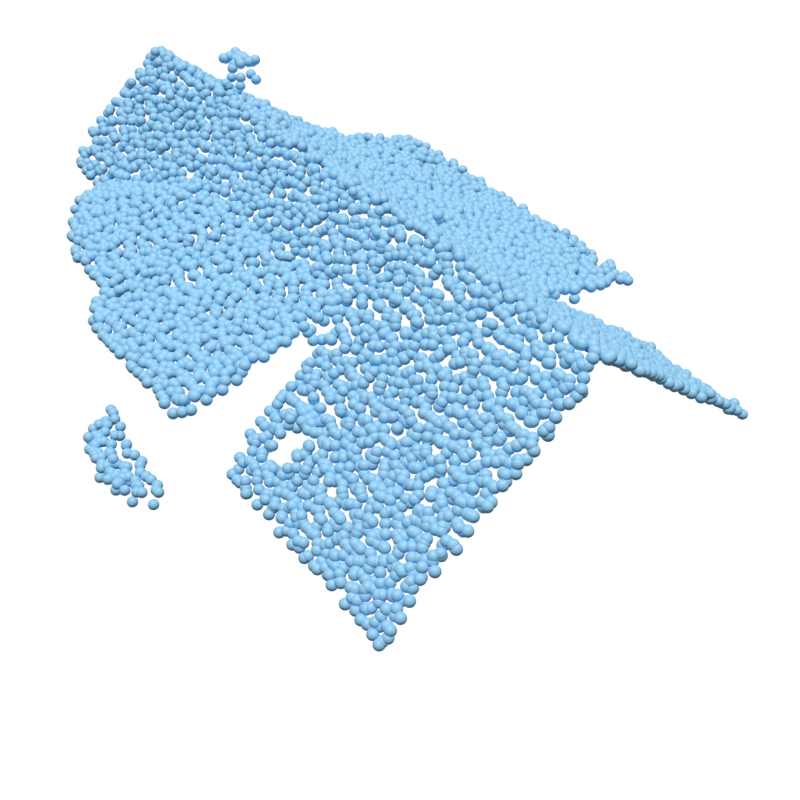}
&
\includegraphics[width=\mywidth,mytrim1]{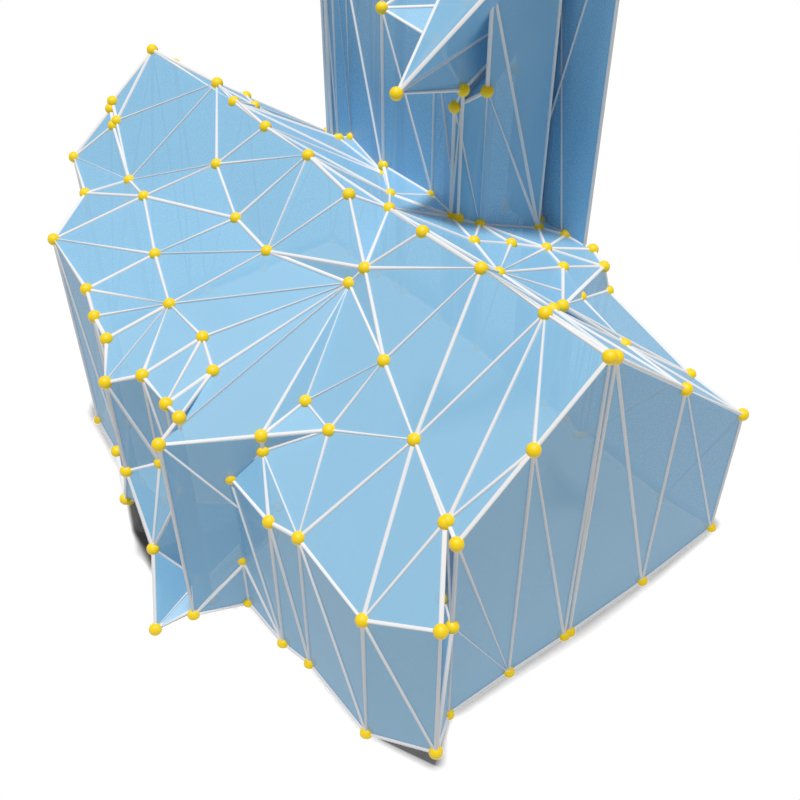}
&
\includegraphics[width=\mywidth,mytrim1]{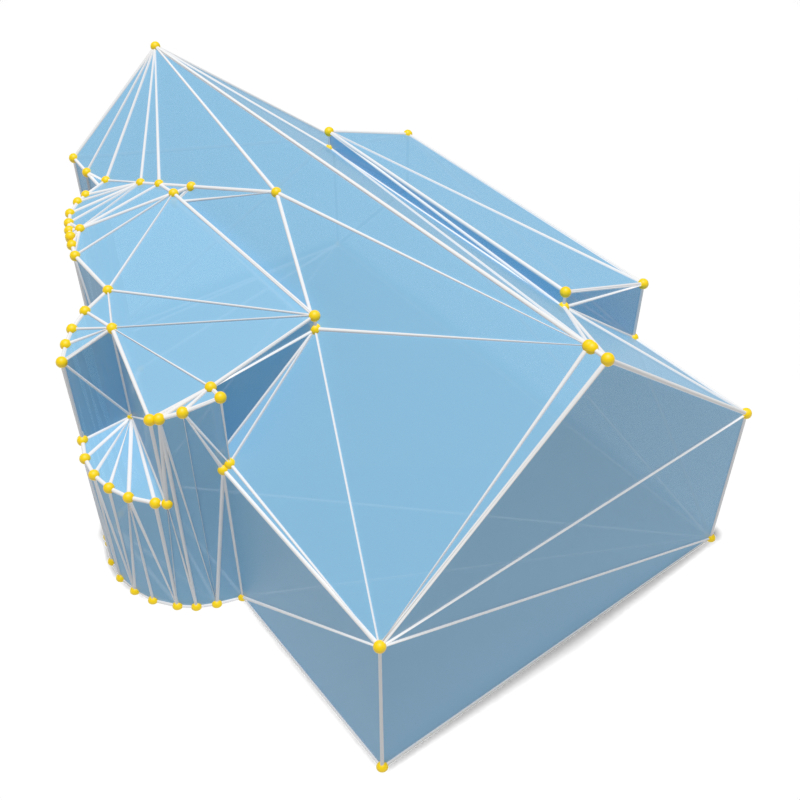}
&
\includegraphics[width=\mywidth,mytrim1]{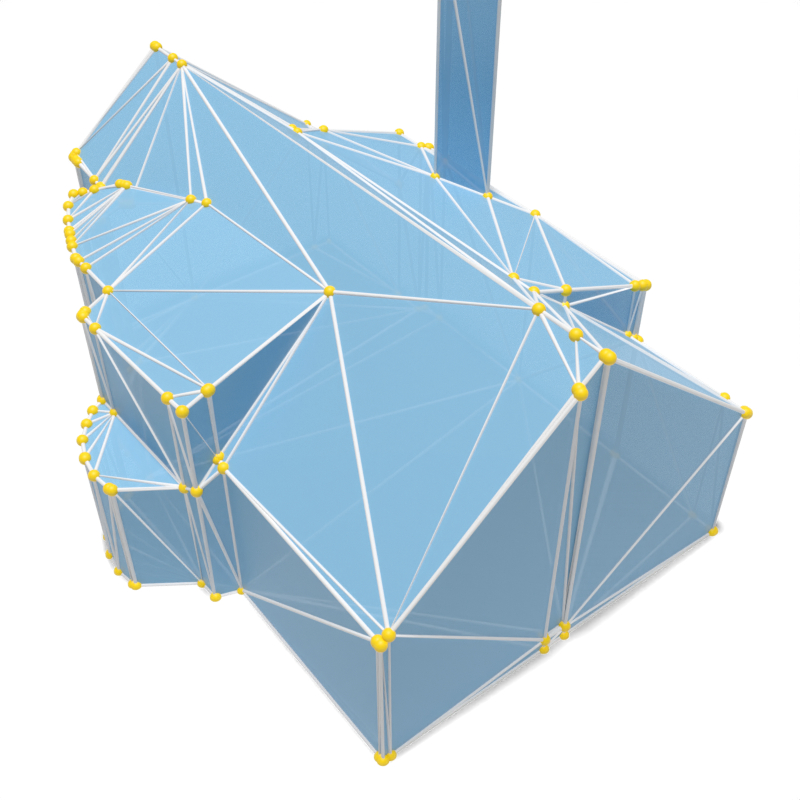}
&
\includegraphics[width=\mywidth,mytrim1]{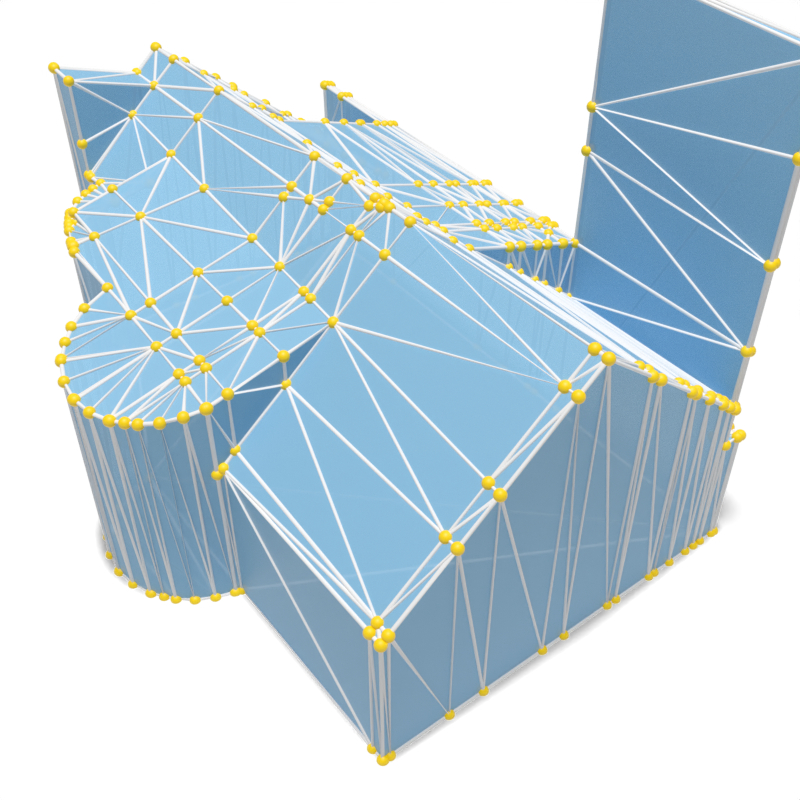}
&
\includegraphics[width=\mywidth,mytrim1]{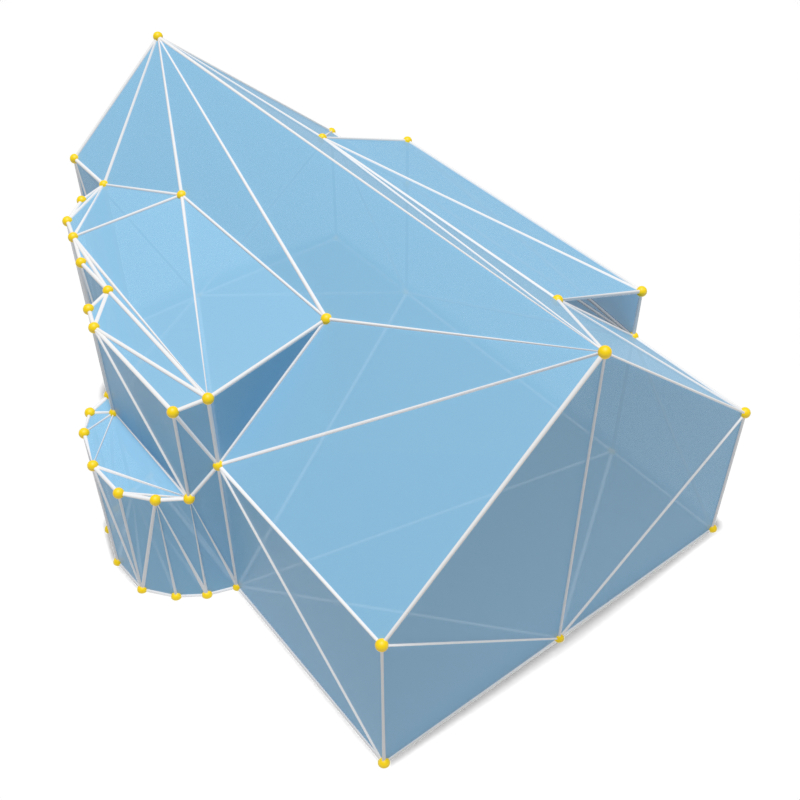}
&
\includegraphics[width=\mywidth,mytrim1]{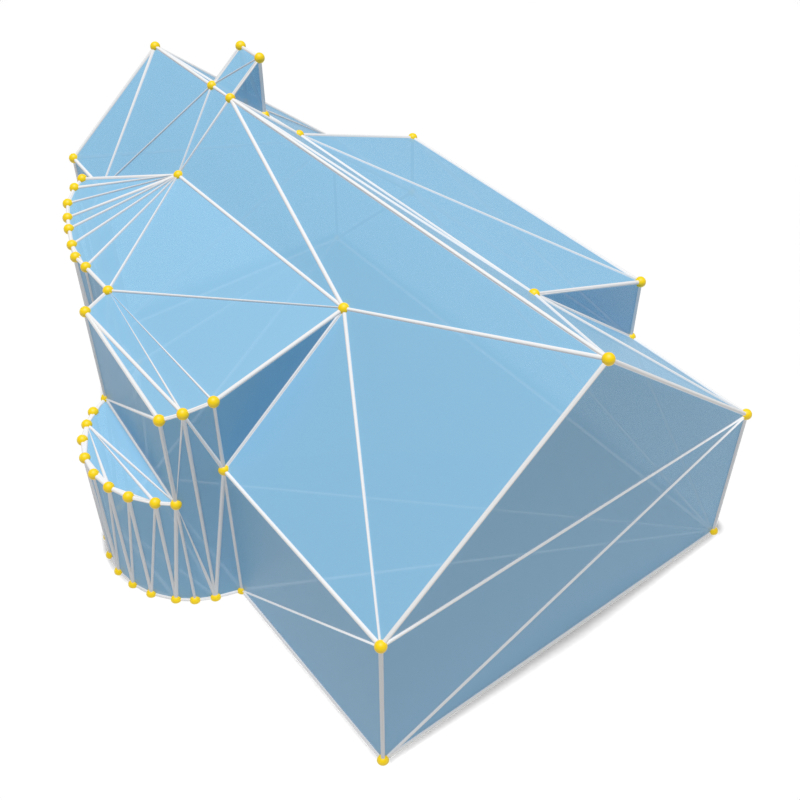}
\\

\multirow{3}{*}{\rotatebox{90}{\hspace{6mm}Zurich}}&
\includegraphics[width=\mywidth,mytrim1]{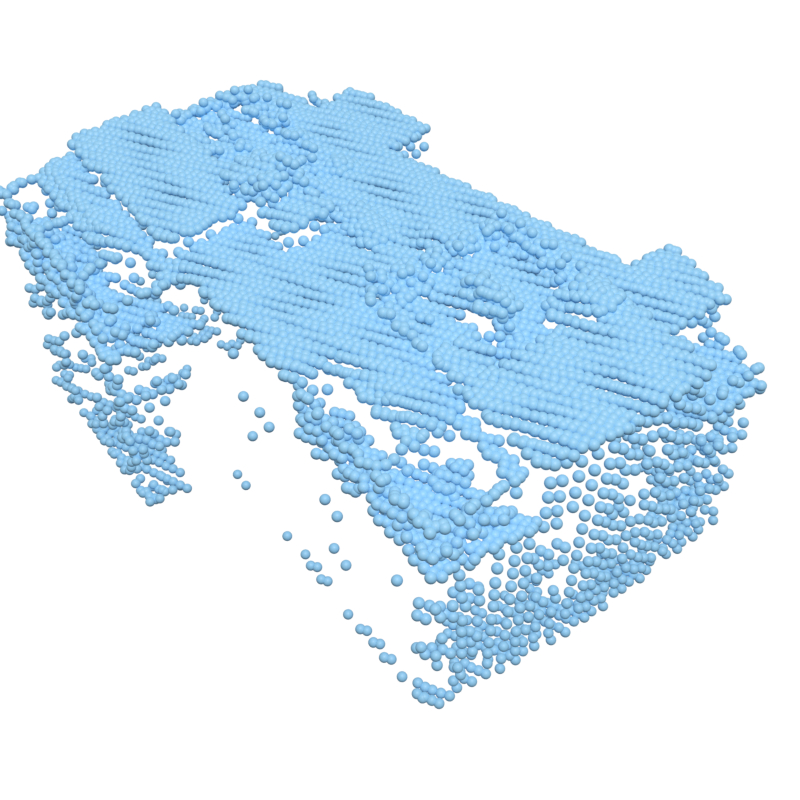}
&
\includegraphics[width=\mywidth,mytrim1]{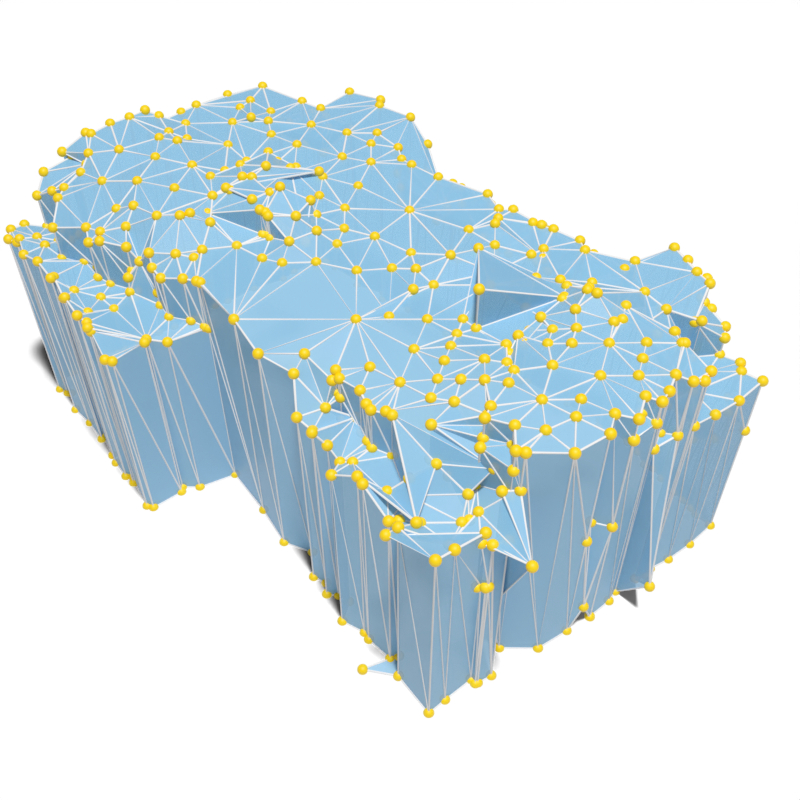}
&
\includegraphics[width=\mywidth,mytrim1]{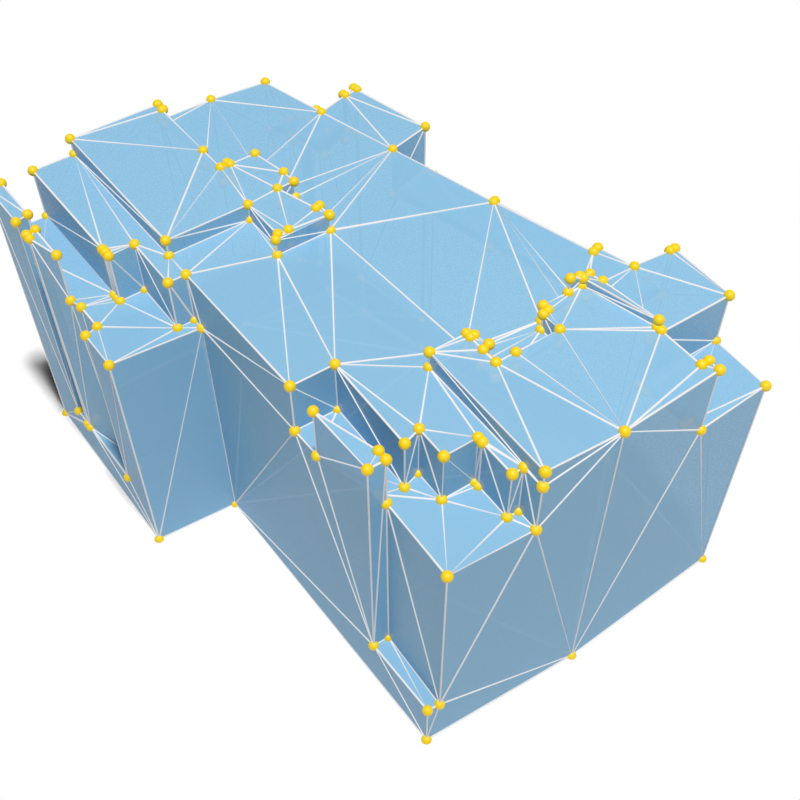}
&
\includegraphics[width=\mywidth,mytrim1]{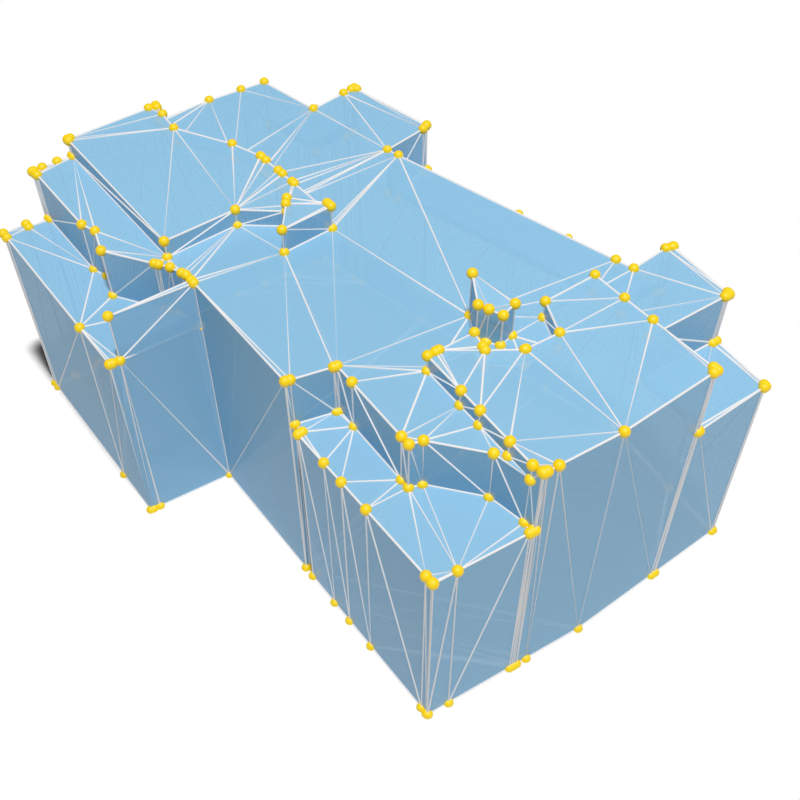}
&
\includegraphics[width=\mywidth,mytrim1]{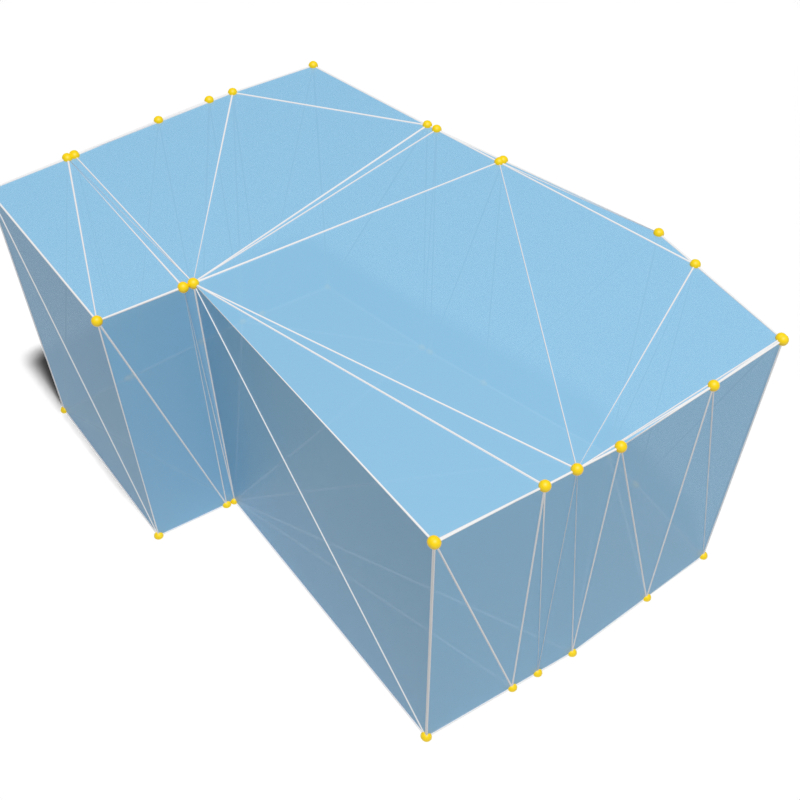}
&
\includegraphics[width=\mywidth,mytrim1]{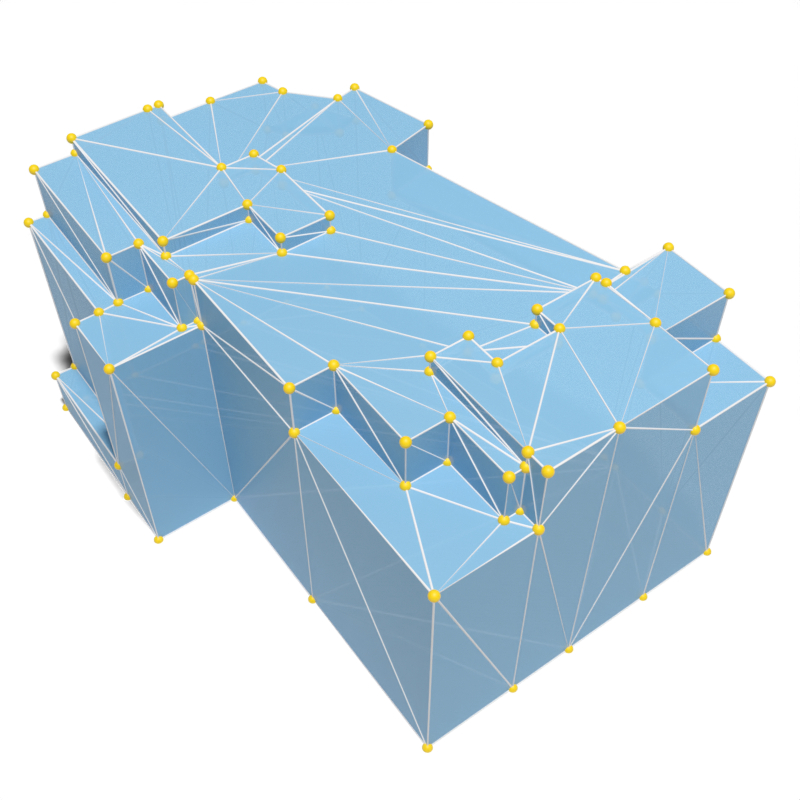}
&
\includegraphics[width=\mywidth,mytrim1]{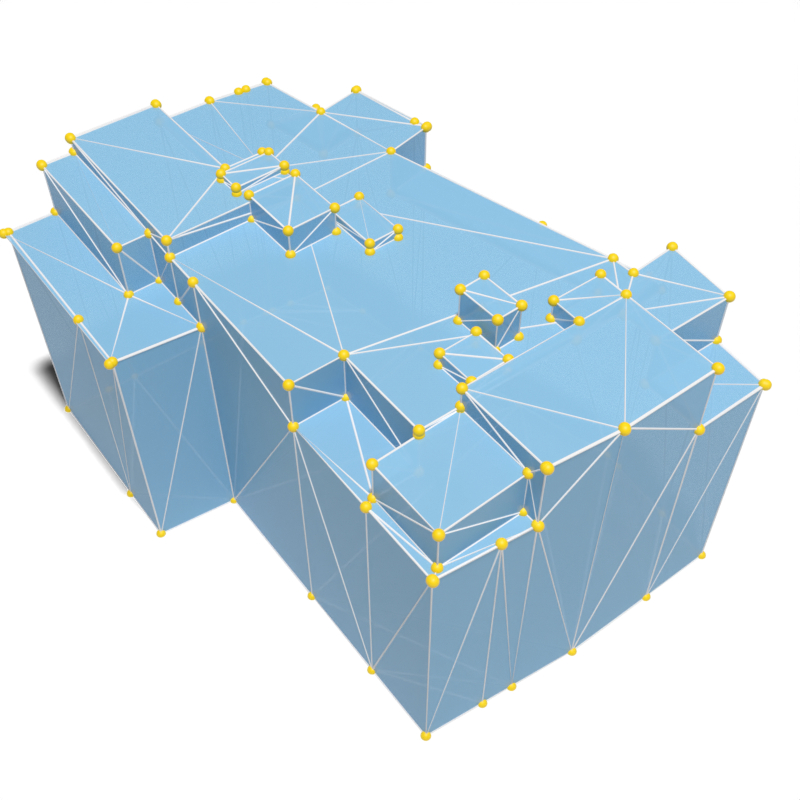}
\\
&
\includegraphics[width=\mywidth,mytrim1]{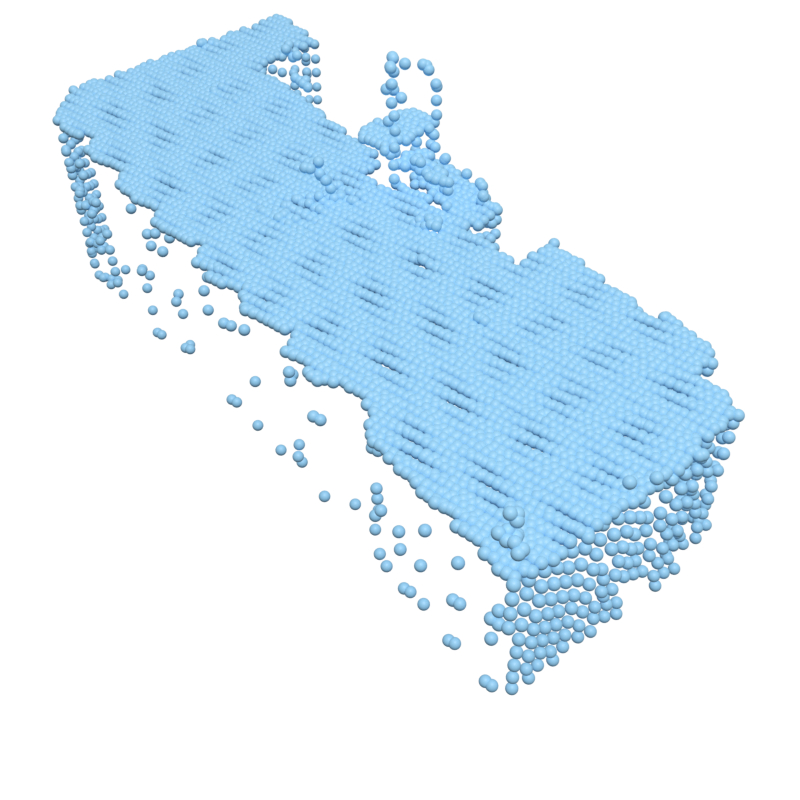}
&
\includegraphics[width=\mywidth,mytrim1]{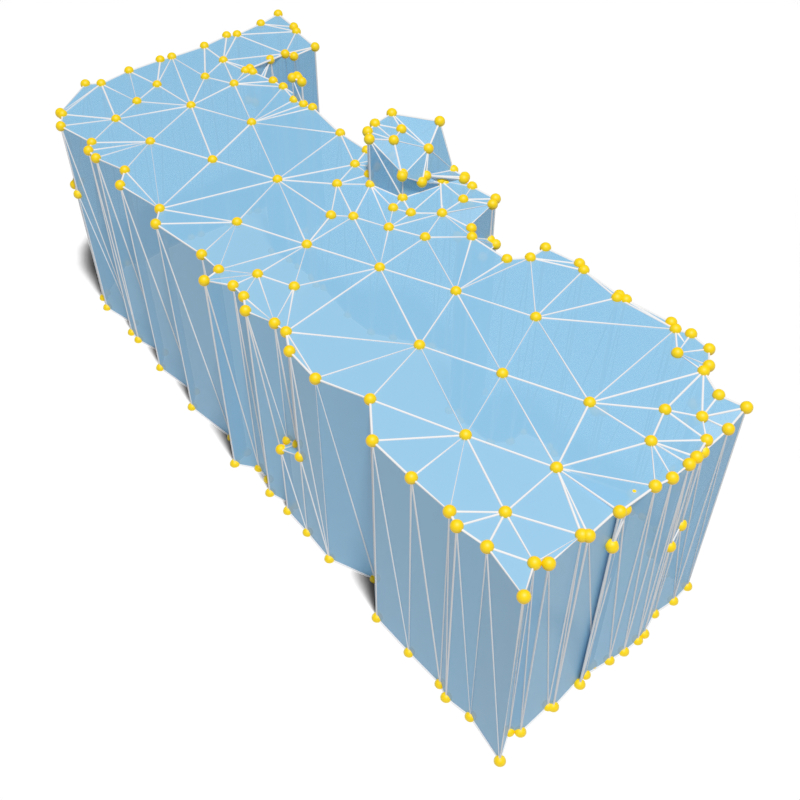}
&
\includegraphics[width=\mywidth,mytrim1]{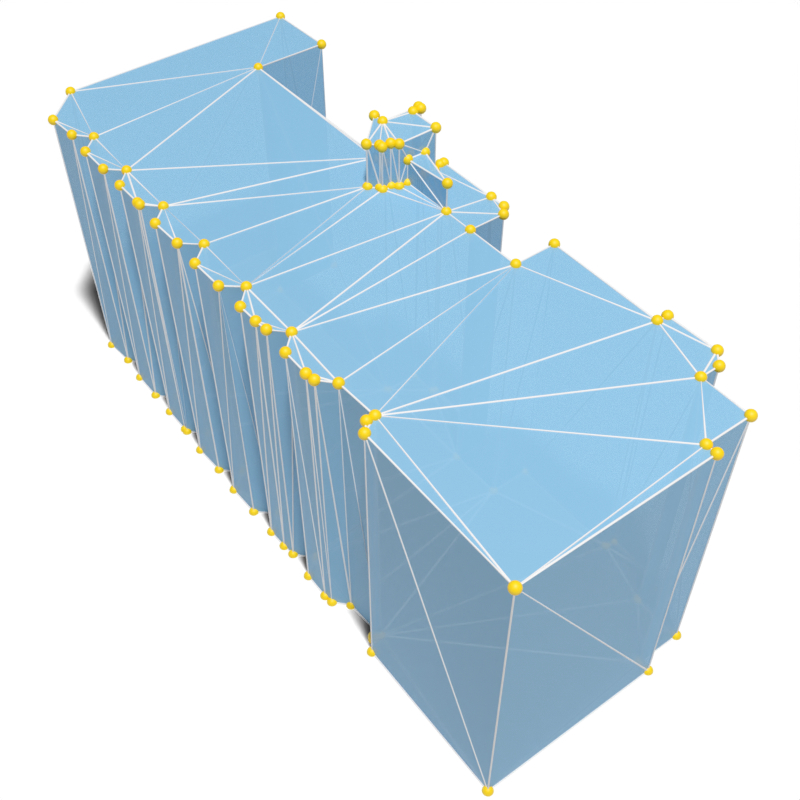}
&
\includegraphics[width=\mywidth,mytrim1]{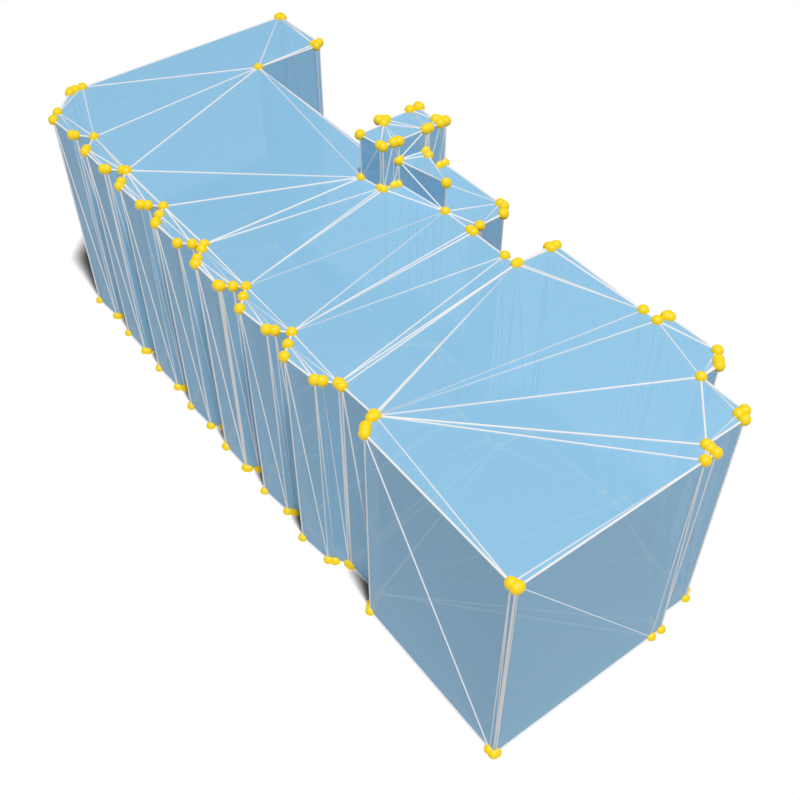}
&
\includegraphics[width=\mywidth,mytrim1]{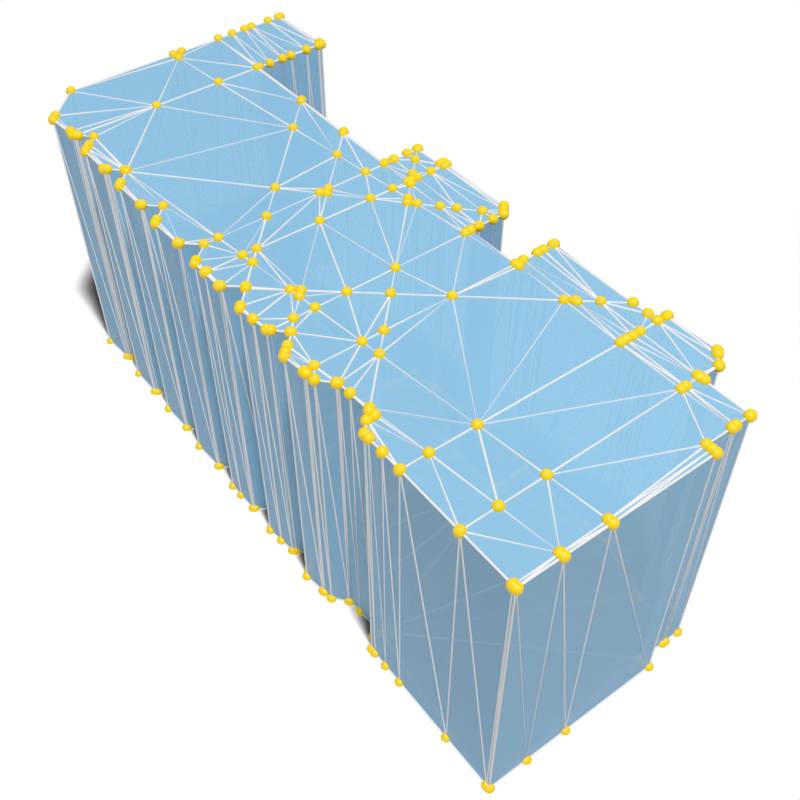}
&
\includegraphics[width=\mywidth,mytrim1]{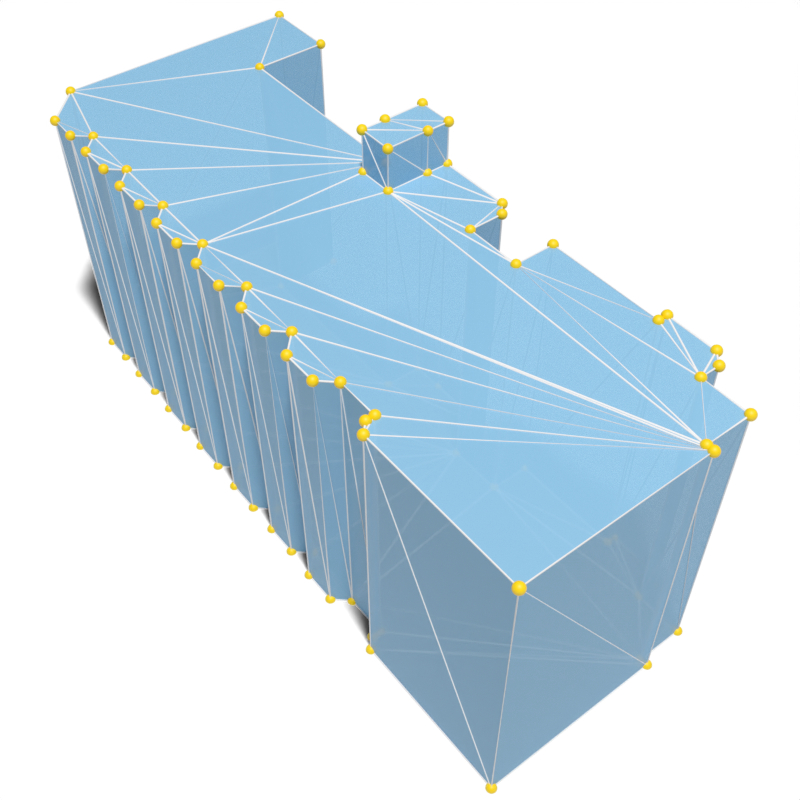}
&
\includegraphics[width=\mywidth,mytrim1]{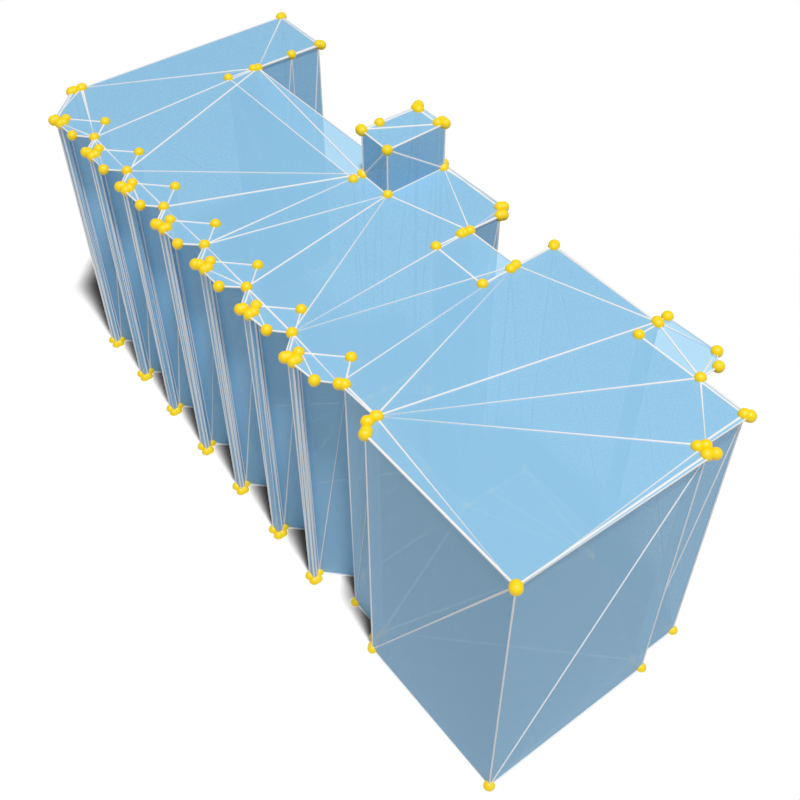}
\\

\multirow{3}{*}{\rotatebox{90}{\hspace{3mm}Helsinki}}&
\includegraphics[width=\mywidth,mytrim1]{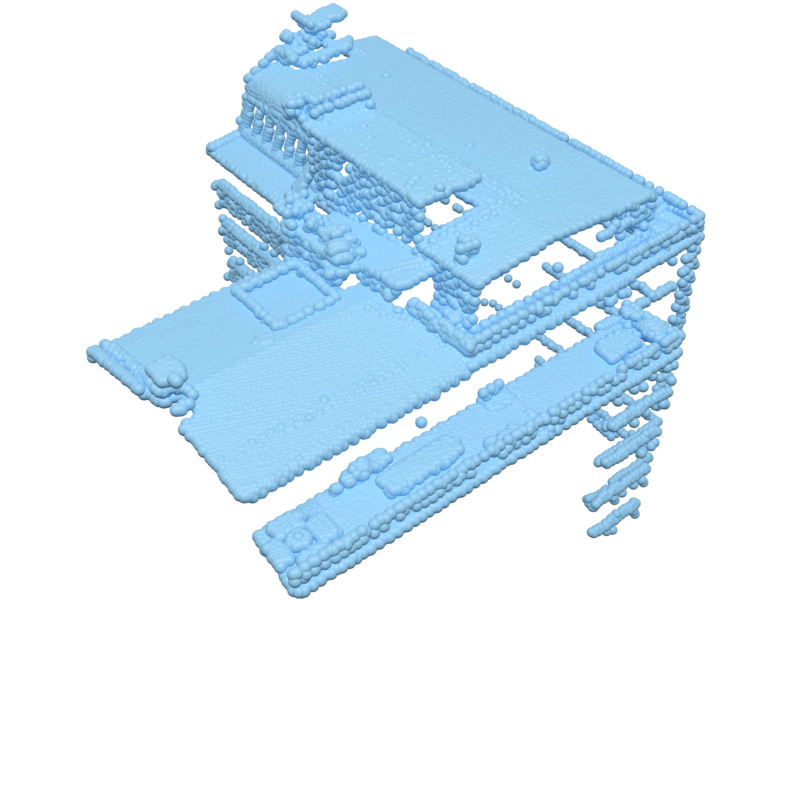}
&
\includegraphics[width=\mywidth,mytrim1]{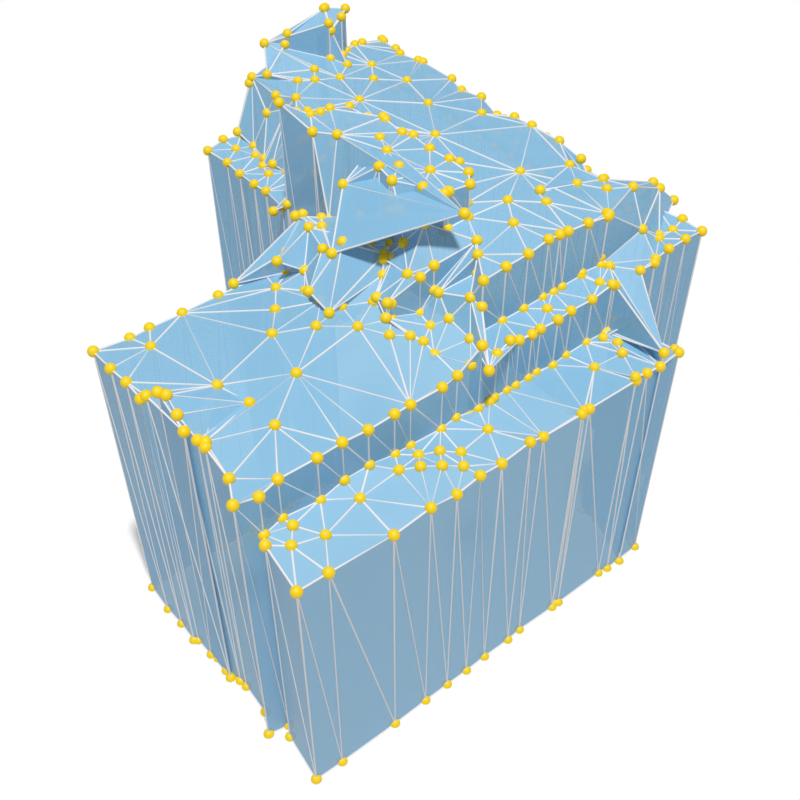}
&
\includegraphics[width=\mywidth,mytrim1]{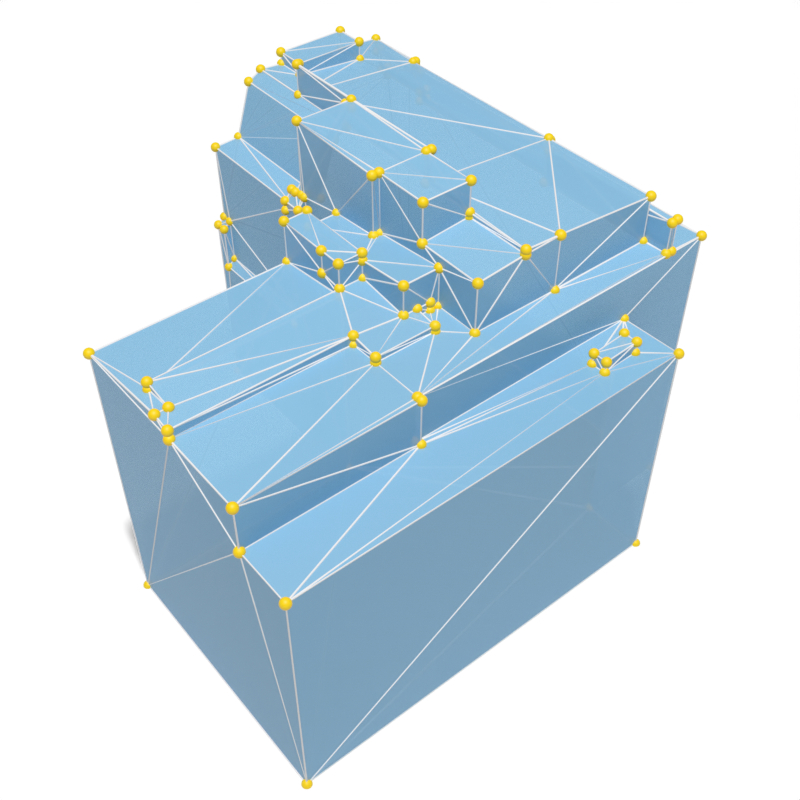}
&
\includegraphics[width=\mywidth,mytrim1]{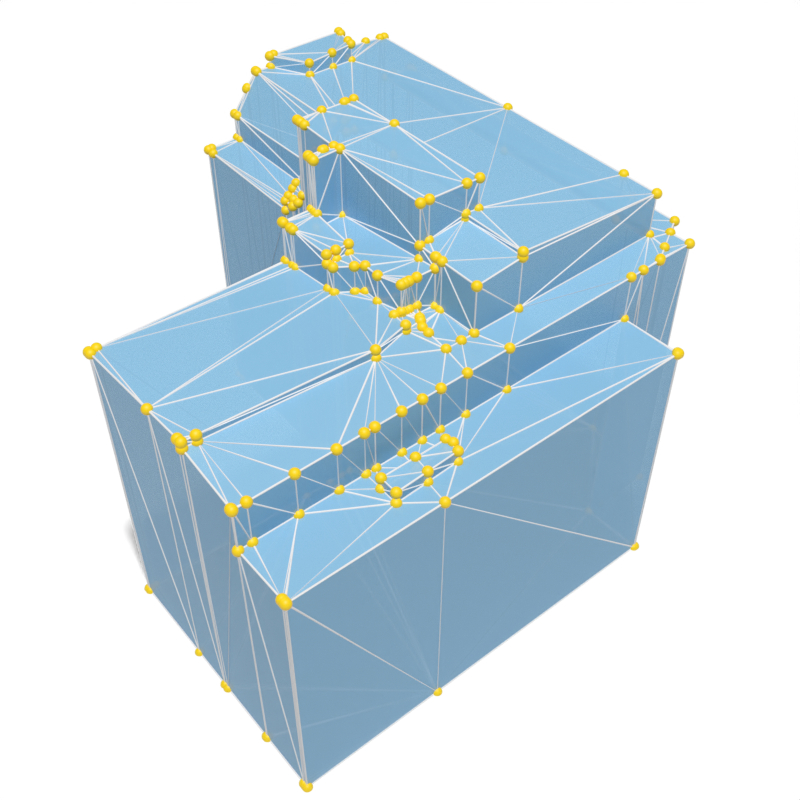}
&
$> 5~\text{min}$
&
\includegraphics[width=\mywidth,mytrim1]{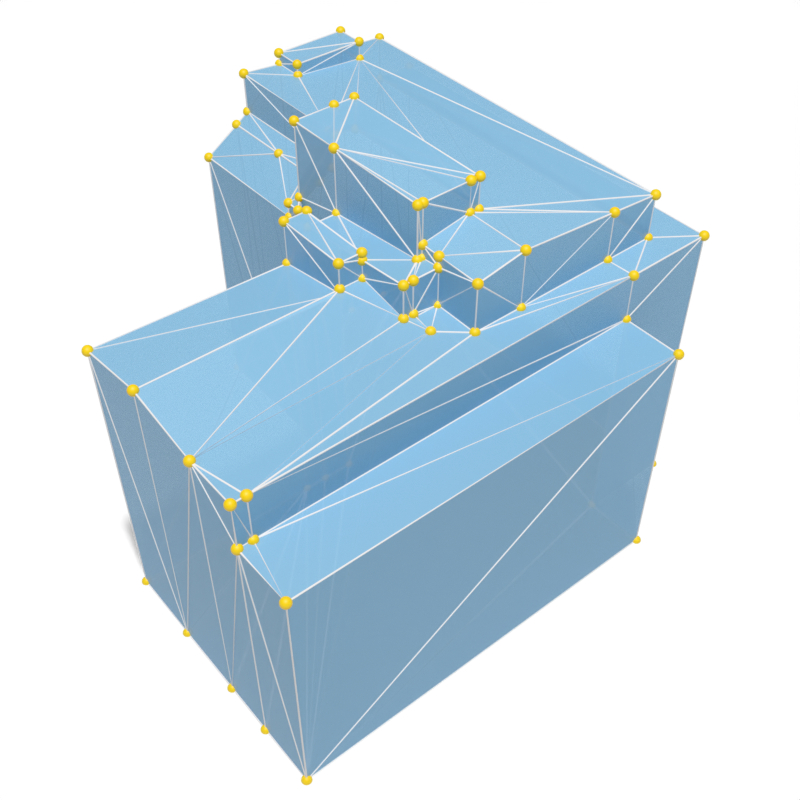}
&
\includegraphics[width=\mywidth,mytrim1]{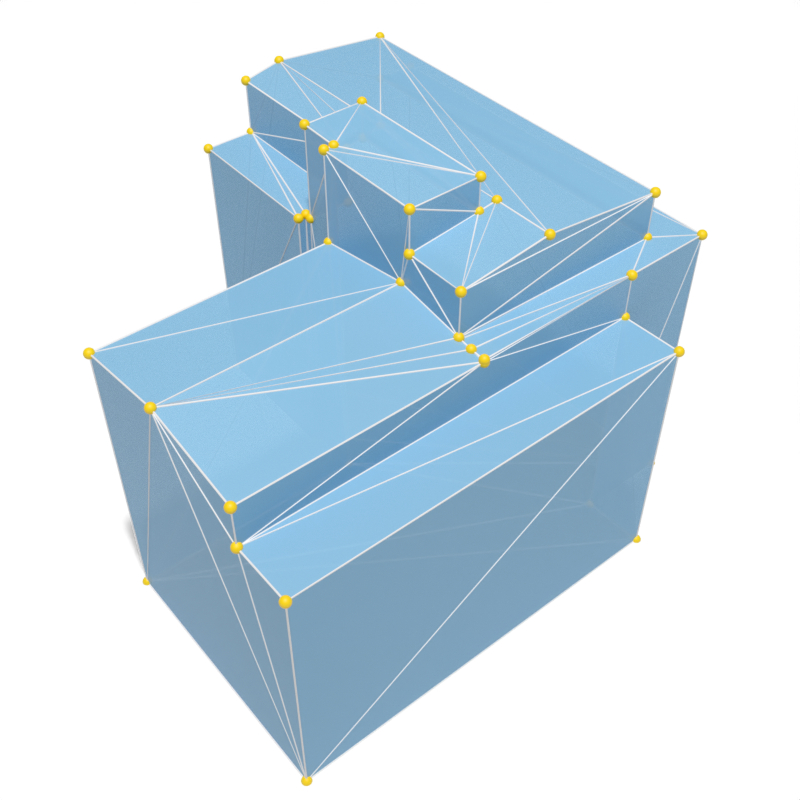}
\\
&
\includegraphics[width=\mywidth,mytrim1]{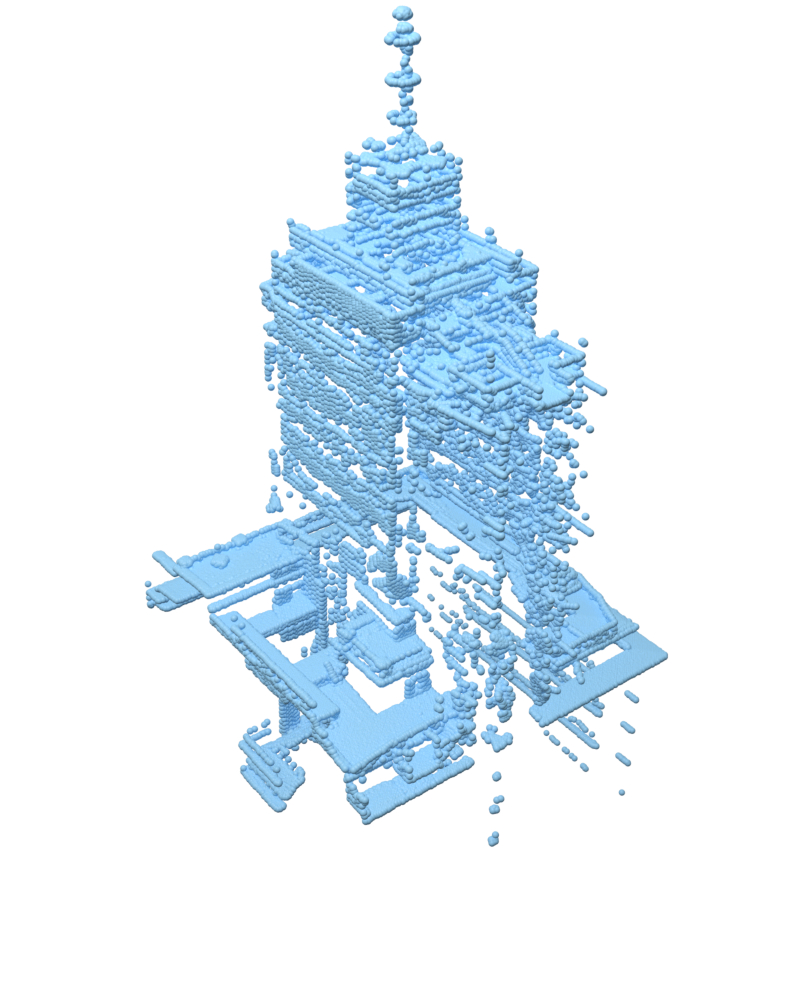}
&
\includegraphics[width=\mywidth,mytrim1]{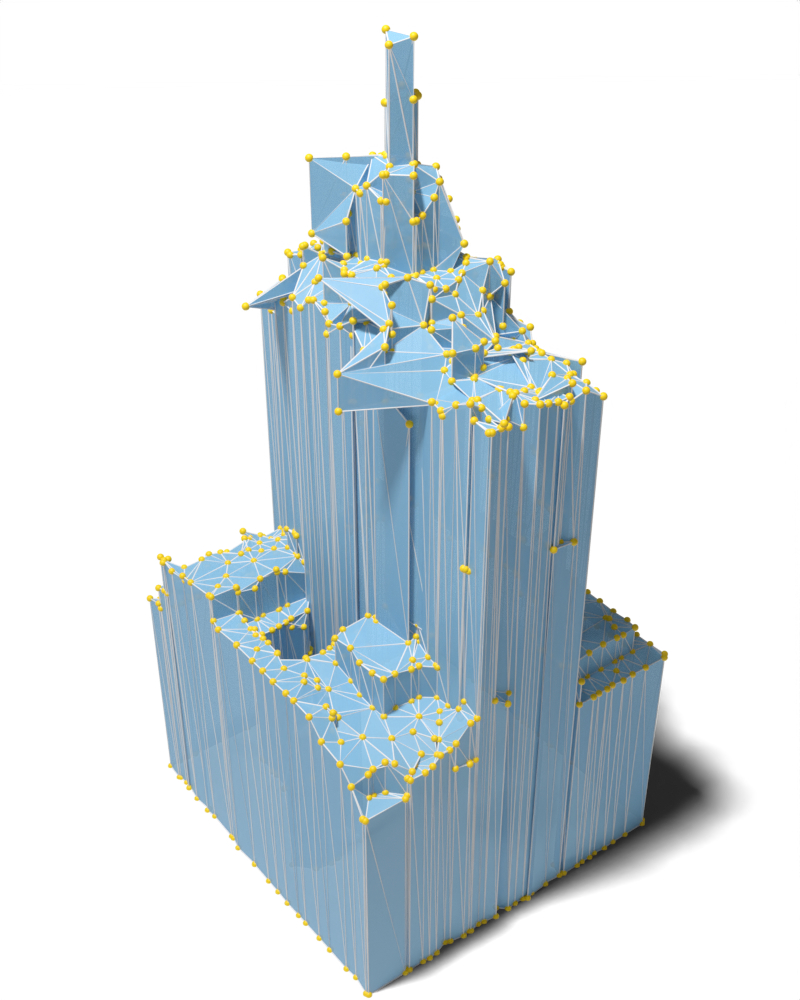}
&
\includegraphics[width=\mywidth,mytrim1]{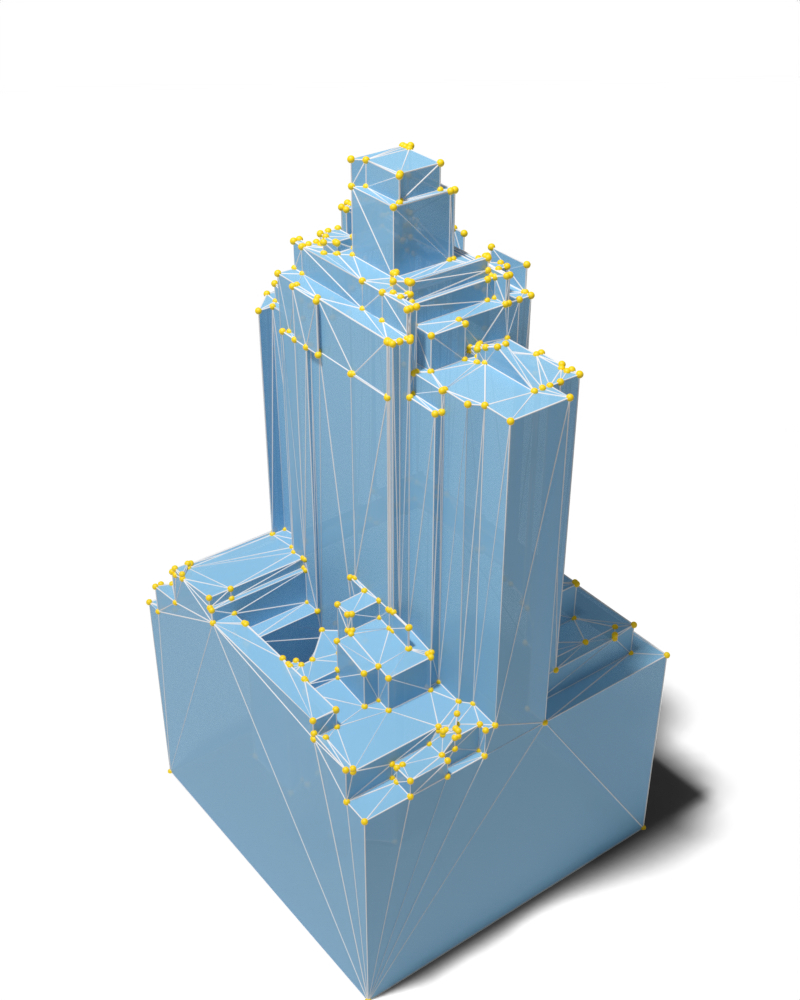}
&
\includegraphics[width=\mywidth,mytrim1]{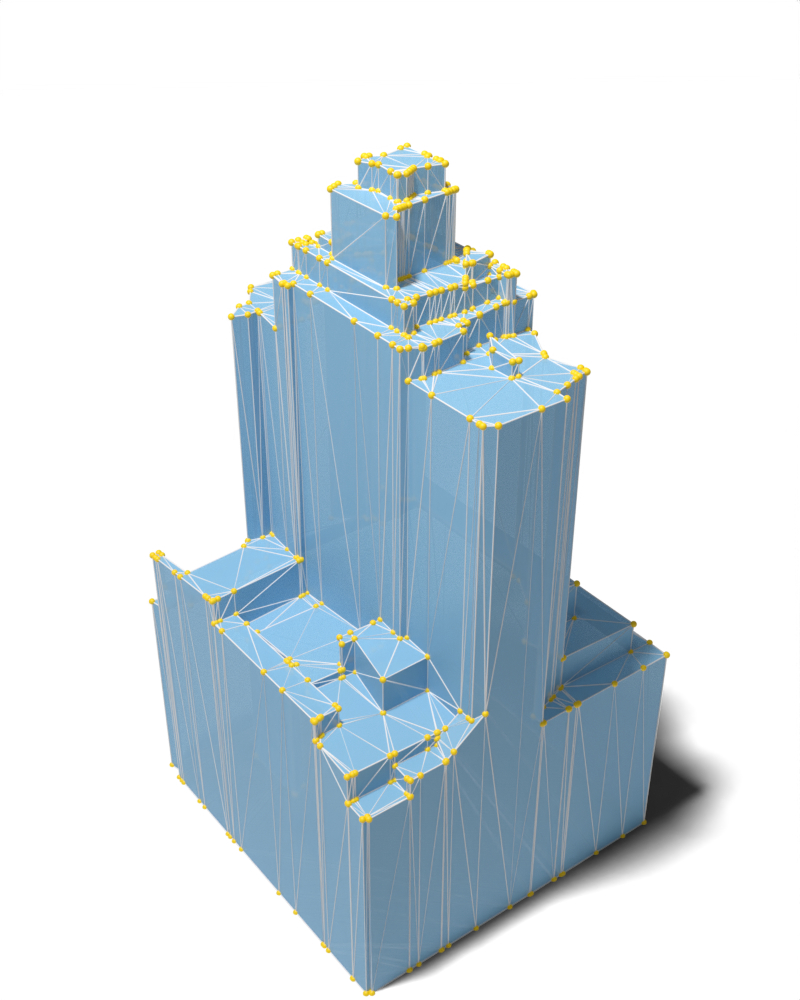}
&
$> 5~\text{min}$
&
\includegraphics[width=\mywidth,mytrim1]{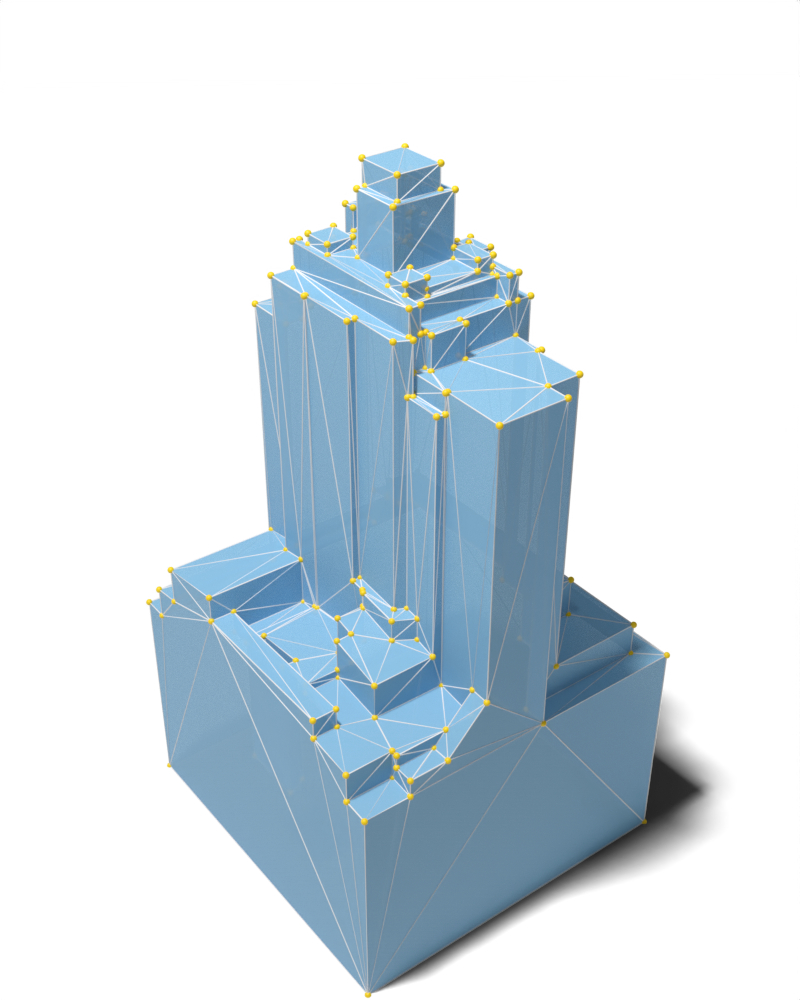}
&
\includegraphics[width=\mywidth,mytrim1]{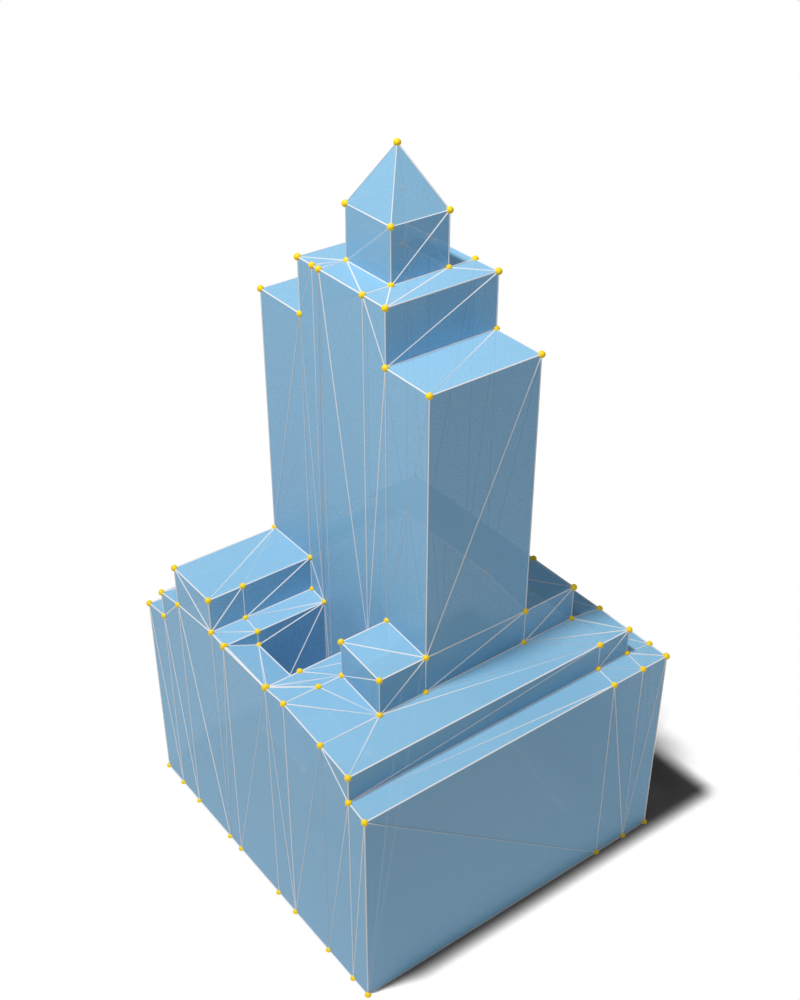}
\\

&
Input & 2.5DC & KSR & Geoflow & City3D & Ours & Reference

\end{tabular}
\caption{\textbf{Visual comparison.} The meshes produced by our algorithm capture the various roof components, even on complex buildings while being simple and regular. In contrast, meshes returned by prior methods are overly complex with respect to the reference models. KSR, Geoflow and City3D generate numerous tiny facets in between the roof sections while 2.5DC suffers from the presence of many visual artifacts. The two top buildings (respectively the third and fourth top buildings, and the two bottom buildings) originate from the \textit{Building3D Tallinn dataset} (respectively the \textit{Zurich dataset} and the \textit{Helsinki dataset}).}
\label{fig:qualitative}
\end{figure*}

\tabref{tab:tallinn} provides the quantitative results of the comparative study while \figref{fig:qualitative} shows visual results on various buildings.

Our method produces meshes that are significantly simpler than competitors on the three datasets. The average number of vertices of the prior methods is two to four times more than ours. This gap is particularly important with Geoflow~\cite{peters2022geoflow} which also exploits a planimetric arrangement-based framework like us, but with a naive extrusion and no regularization of the partition. Our meshes also exhibit a much lower ratio of small edges than the competitors, except for 2.5DC \cite{zhou2010dualcontouring} that, by construction, displaces hermite vertices, but without guaranteeing the planarity of roof sections and facade components.

The significant gain on complexity metrics does not affect performance as our method is among the fastest behind 2.5DC \cite{zhou2010dualcontouring} and more than one order magnitude faster than City3D \cite{huang2022city3d}. Notably, reducing complexity also limits expensive downstream operations on the geometric elements forming the output mesh. The gain in regularity moderately impacts accuracy with typically a precision loss of a few centimeters with respect to the most accurate methods. Note that our method exhibits a competitive score on both Chamfer distances (i.e. measured from input points and to reference model), in contrast to 2.5DC \cite{zhou2010dualcontouring} or Geoflow~\cite{peters2022geoflow} that perform well on only one of the two.

Regarding the geometric properties of the reconstructed models, our algorithm offers the most desired guarantees with watertight, 2-manifold, intersection-free meshes. Similarly to 2.5DC \cite{zhou2010dualcontouring} and Geoflow~\cite{peters2022geoflow}, our algorithm relies upon a 2.5D-view dependent representation of buildings. Only KSR \cite{bauchet2020kinetic} and City3D \cite{huang2022city3d} can produce full 3D models, but also require higher computing resources.

\begin{figure*}[!ht]
\centering
\begin{subfigure}{0.49\textwidth}
\includegraphics[width=\columnwidth]{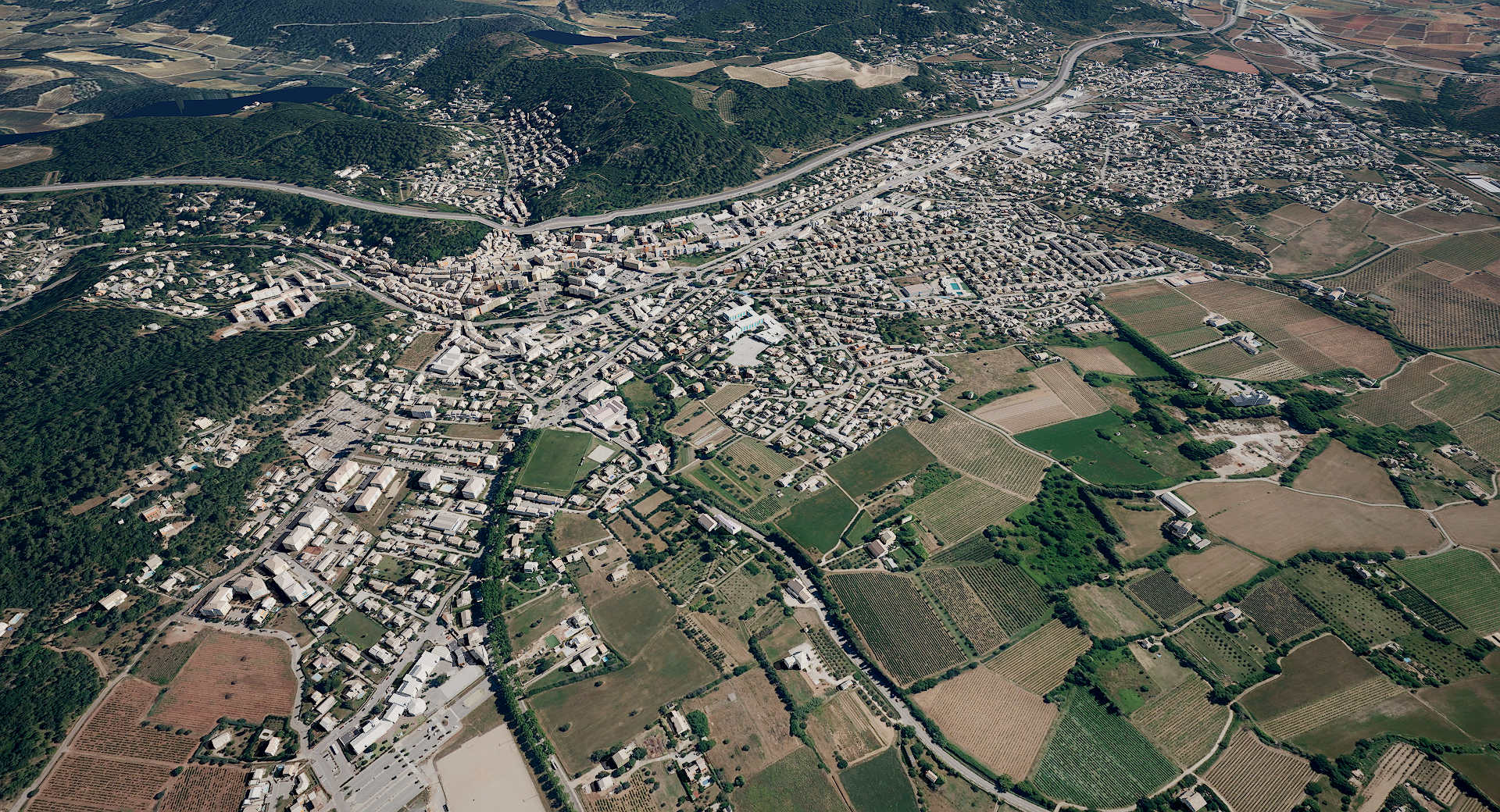}
\end{subfigure}
\hfill
\begin{subfigure}{0.49\textwidth}
\includegraphics[width=\columnwidth]{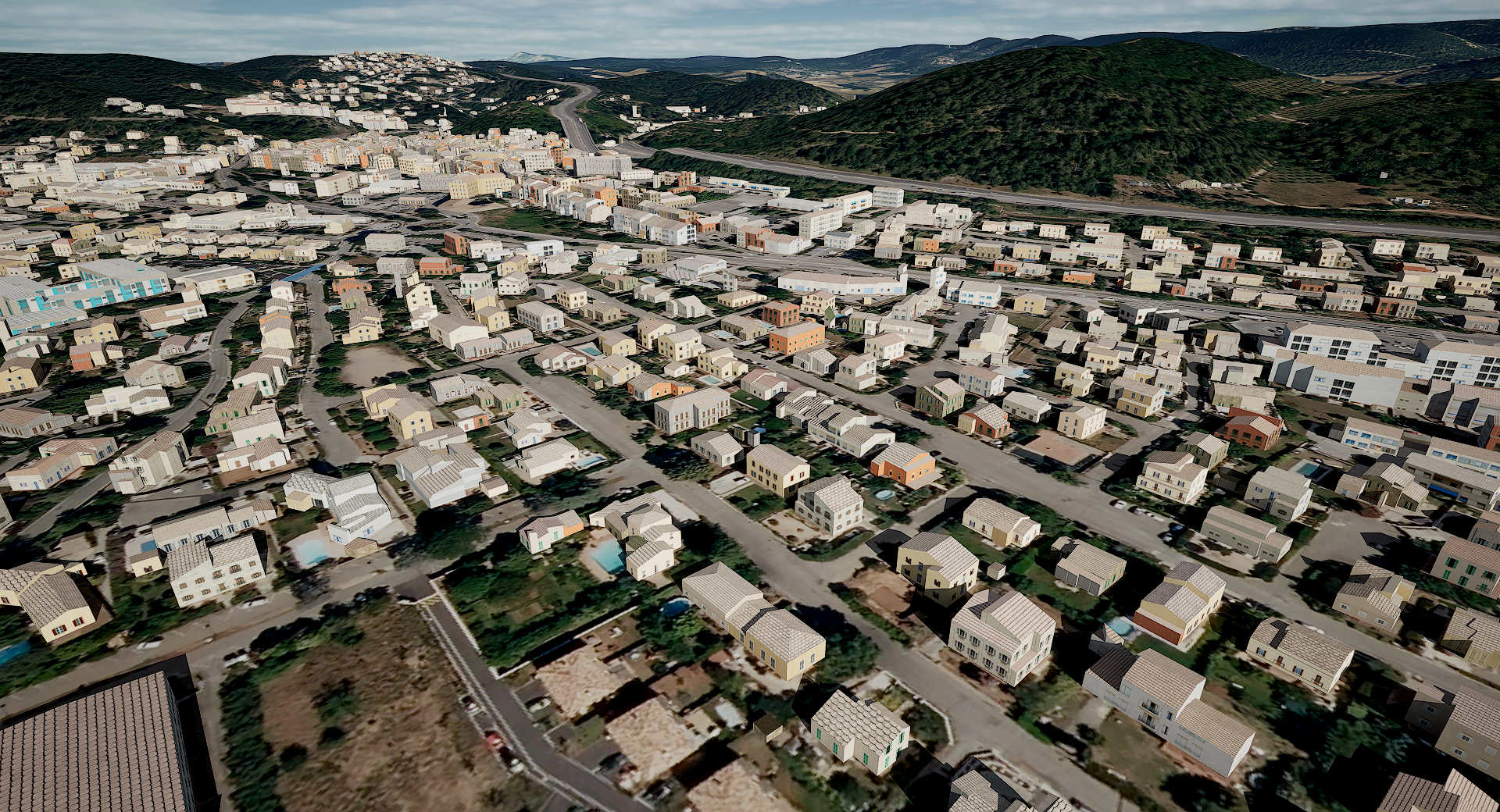}
\end{subfigure}
\caption{\textbf{Textured reconstructions of Le Luc, France.} Both airborne Lidar point clouds and cadastral maps available online \cite{ign} were used as input to generate 3D building models, prior to the texturing process. Best viewed in electronic version.}
\label{fig:textured}
\end{figure*}

\subsection{Ablation}

To validate our pipeline we ablate the main design choices and reconstruct a subset of the Tallinn building models. We show the results in \tabref{tab:ablation}. We first remove the discontinuity lines from the input of the 2D polygonal partition. This results in a simpler partition without the inner vertical discontinuities, and finally in models with slightly fewer vertices and facets, but with an increased average distance from the input point cloud of around 4~cm. 
Conversely, disabling the two optimization procedures of Sections \ref{sec-step2} and \ref{sec-step3}, or one of them only, leads to more complex models that satisfy less geometric guarantees but exhibit a slightly better accuracy, \ie less than 1~cm on average on the two Chamfer distances.


\begin{table}[t]
    \setlength{\tabcolsep}{9pt}
\centering
\caption{\textbf{Ablation study.} We ablate the main design choices of our pipeline (row 1) and evaluate reconstructions over a subset of 1,000 buildings of the Tallinn dataset. 
We remove vertical discontinuities from the construction of the 2D kinetic partition (row 2), both regularization steps (row 3), or apply only the vertical (row 4) or horizontal (row 5) regularization.
}
\vspace{-0.2cm}
\resizebox{\linewidth}{!}{%
\begin{tabular}{@{}lrrrrr@{}}
\toprule
& \multicolumn{3}{c}{\emph{Complexity}} & \multicolumn{2}{c}{\emph{Accuracy}}   \\
\midrule
                                                    & $|V|$         & $|F|$         & $E_{<50~\text{cm}}$  &  $\recall$&  $\precision$         \\
&   &  &  [\%] & [cm] & [cm]   \\
\midrule
\textbf{Ours}                                    & \ts 32.7      & \ts 58.3       & \ts 5.01             &  9.11     & 12.8             \\
\textbf{~No vertical discontinuities}               & \tf 29.5      & \tf 53.1       & \tf 4.89             &     13.5  & \tf  10.5          \\
\textbf{~No regularization}                         &    38.7       & 66.7           &    8.13              & \ts 8.13  &  12.6             \\
\textbf{~Only vertical reg.}                        &    34.2       & 62.1           &    6.41              & 8.54      &  \ts 12.4          \\
\textbf{~Only horizontal reg.}                      &    34.0       & 59.9           &      5.3             & \tf 8.02   &      13.1         \\
\bottomrule
\end{tabular}
}
\label{tab:ablation}
\end{table}

\subsection{Application to city texturing}
\label{application}
The ability of our algorithm to produce simple and regularized meshes
is particularly interesting for simulation tasks, but not only. It is also valuable for visualization scenarios. 


To this end, we adapt the inverse procedural modeling pipeline of Girard \etal \cite{girard2023brightearth}.
Instead of modeling building geometries using remote imagery and shape grammars, we directly use 3D models generated by our pipeline as input of the texturation pipeline.
Roofs are then textured based on their shape (flat, gabled, hipped, etc) using a library of predefined high-resolution textures, and ground-based imagery is used to extract the most similar element from a library of high-resolution facade textures, to preserve the architectural style of the building.
Fig.~\ref{fig:textured} shows the result of such texturing process at a city scale.


\subsection{Limitations}

Our algorithm has a few limitations. Firstly, the 2.5D modelling approach yields LOD2.2 models and does not allow to reconstruct 3D building features such as roof overhangs and balconies (which would correspond to LOD2.3), that are yet present in many real-world cases. Besides, the robustness of our method relies upon the quality of the detected plane configurations. In particular, if the plane detection technique misses one important planar element, then our reconstruction pipeline will be unable to reconstruct the related wall or roof section. This can make some near-vertical structures, such as steeples, hard to reconstruct, because they are represented by a limited number of points in airborne Lidar data. Curved structures, such as domes, are also approximated by a piecewise-planar geometry. Finally, our reconstruction pipeline assumes the input building footprints and Lidar scans align well. Misalignments problems are however frequent, especially when using online cadastral maps. 

\section{Conclusion}
\label{conclusion}

We present SimpliCity, a building reconstruction method that produces simple, regularized 3D models while offering a similar efficiency and fidelity to input data compared to prior pipelines. Our planimetric arrangement-based framework relies upon two key ideas: (i) local and global regularization of a 2D polygonal partition built from detected 3D planes, and (ii) extrusion of the partition with guaranteed planarity of roof sections and preservation of vertical discontinuities and horizontal rooftop edges. 

In the future we would like to extend our optimization techniques to produce regularized LOD3 building models. Besides, we intend to investigate on the building segmentation and contouring problems, to correct issues induced by incorrect or incomplete building footprints. We also plan to release our reconstruction pipeline as a web service.

\section*{Acknowledgements}

This research work was funded by the DGA GENESE project. The authors warmly thank Vincent Madelain for fruitful discussions, and Cédric Larrosa, Enora Fiker, Yacine Kacem for the proposed help for conducting experiments.

\FloatBarrier
{
    \small
    \bibliographystyle{ieeenat_fullname}
    \bibliography{main}
}


\end{document}